\documentclass[12pt,a4paper]{article}
\pdfoutput=1
\usepackage{jheppub}
 \usepackage{amsmath}
 \usepackage{amsfonts}
 \usepackage{mathtools}
 \usepackage{amssymb}
 \usepackage{caption}
 \usepackage{subcaption}
 \usepackage{slashed}
 
 \title{Spectral sum rules for conformal field theories in arbitrary dimensions}
 \author{Subham Dutta Chowdhury ${}^a$, Justin R. David ${}^a$, Shiroman Prakash ${}^b$}
\affiliation{ ${}^a$ Centre for High Energy Physics, Indian Institute of Science,\\
C. V. Raman Avenue, Bangalore 560012, India.\\
${}^b$Department of Physics and Computer Science, \\
Dayalbagh Educational Institute, Dayalbagh, \\
Agra 282005, India.}
\emailAdd{subham, justin@cts.iisc.ernet.in, shiroman@gmail.com}
\abstract{We derive spectral sum rules in the shear channel for
conformal field theories at finite temperature  in general $d\geq 3$ dimensions.
The sum rules result from the OPE of the stress tensor at high
frequency as well as the hydrodynamic behaviour of the
theory at low frequencies.
The sum rule states that a weighted integral of the spectral
density over frequencies is proportional to the energy density of the theory. 
We show that the proportionality constant can be written in terms
the Hofman-Maldacena variables $t_2, t_4$ which determine 
 the three point function of the
stress tensor. 
For theories which admit a two  derivative gravity dual
this  proportionality constant is given by  $\frac{d}{2(d+1)}$  .
We then use causality    constraints  and
obtain bounds on the sum rule which are valid in  any conformal field theory.
Finally  we  demonstrate that the high frequency behaviour of the
spectral function in the vector and the tensor channel
are also determined by  the Hofman-Maldacena variables.}
\begin{document}
\maketitle
\section{Introduction}

 Sum rules for spectral densities in any quantum field theory provide 
 important data of the theory. 
 The sum rule relates  a weighted integral of the spectral density 
 over  frequencies to one point functions of the theory. 
They  result  because of the    analyticity of the corresponding Greens function 
  together with the short distance as well as the long distance 
  behaviour   of the theory. Thus sum rules  provide useful constraints on 
  spectral densities.
  For instance  real time finite temperature retarded correlators
  are difficult to obtain from lattice calculations in QCD. However 
  one point functions are considerably easier to obtain. This has led to 
  a systematic study of the sum rules which constrain the 
  spectral densities of the stress tensor in  QCD 
  \cite{Kharzeev:2007wb,Karsch:2007jc,Romatschke:2009ng,Meyer:2010ii,Meyer:2010gu}. 
  Similarly there are sum rules studied in condensed matter
   like the Ferrell-Golver-Tinkham sum rule 
  which is satisfied by the current-current correlator 
  in a BCS superconductor \cite{Ferrell:1958zza,PhysRevLett.2.331}
  or the sum rule  for momentum distribution in 
  angle resolved photon emission  \cite{PhysRevLett.74.4951}.

Schematically  the sum rules we will focus on have the structure 
 \begin{eqnarray}\label{bassum}
 \int_{-\infty}^{\infty} \frac{d\omega}{\omega^n} \rho(\omega) &\propto& 
  \langle \textrm{One point functions} \rangle, \nonumber\\
 \rho (\omega) &=& \textrm{Im} G_R(\omega),
 \end{eqnarray}
where $G_R(\omega)$ is the retarded Green's function at finite temperature and 
zero momentum.   
As we have mentioned, 
sum rules are the result of 
 the analytic properties of the Green's function in the upper half plane
 which in turn follows    from causality. 
 Although sum rules primarily relate two point functions to one point functions, we will see that they contain information about the three point functions also. 
 This information will be contained in the proportionality constant 
 in (\ref{bassum}). 
 
 In order to emphasise the usefulness of such rules, 
 let us recall how such sum rules are used to determine 
 transport  properties of quark gluon plasma from the lattice. 
 Here one postulates the form of the spectral density and then fits the 
 parameters involved in the postulated form of the spectral density  from 
 lattice data \cite{Nakamura:2004sy,Aarts:2007wj,Meyer:2007ic,Meyer:2007dy,Huebner:2008as}. 
 Since sum rules are constraints satisfied by the spectral density, they 
 restrict the class of the postulated form for the spectral density. 
 For example the simplest Lorentzian anstaz for the 
 spectral density $\frac{ \omega}{\omega^2 + \Gamma^2}$ is disallowed 
 using the shear sum rules in QCD \cite{Romatschke:2009ng}.

 All the above  properties make sum rules an important object of interest in quantum 
 field theories. 
 In this paper we will derive 
  sum rules corresponding to the spectral density of the retarded correlator of the 
  $T_{xy}$ component of the stress tensor for 
  an arbitrary  conformal field theories in $d \geq 3 $ dimensions. 
  This sum rule is usually referred to as the shear sum rule.  
  In the context of conformal field theories, sum rules can be used 
  to find conditions under which a certain conformal field theory admits 
  a gravity dual. 
 To illustrate this, we use the shear sum rule to find a necessary condition 
  under which  given conformal field theory admits a gravity dual. 
  This condition is independent of the equality of central charges $a=c$ which 
  is known in the literature.  Sum rules together with causality 
  can be used to find constraints both for conformal field theories as well 
  as putative gravity duals. 
  We will obtain these constraints in  section \ref{bound} of this paper.

The investigation of sum rules in conformal field theories and its relation 
with holography was first done in \cite{Romatschke:2009ng}. For 
$\mathcal{N}=4$ supersymmetric Yang-Mills theory at finite temperature the shear 
 sum rule is given by 
\begin{eqnarray}\label{n4sum}
 \frac{1}{\pi} \int_{-\infty}^{\infty}
  \frac{d\omega}{\omega} \left[ \rho (\omega) - \rho_{T=0}(\omega) \right]
 = \frac{2}{5} \epsilon ,
\end{eqnarray} 
where 
\begin{eqnarray}
\rho(\omega)= \textrm{Im} G_R(\omega), 
\end{eqnarray} 
is the spectral density corresponding to the retarded shear correlator defined  in 
position space by 
\begin{equation}
G_R(t, \vec x)=  i\theta(t)\langle [T_{xy}(t, \vec x),T_{xy}(0)] \rangle.
\end{equation}
The expectation value is taken in the theory held at  finite temperature $T$ and 
$\epsilon$ in (\ref{n4sum})  is the energy density in the theory. 
The authors proposed the relation (\ref{n4sum})  from field theory arguments 
and then verified it using a holographic computation of the 
Greens  function in the $AdS_5$ black hole background. 
The subsequent works mainly focused on the holographic derivation of
the sum rules. 
In \cite{Gulotta:2010cu} the shear sum rule was derived  in holography 
by considering a black hole in $AdS_{d+1}$ dimensions. 
Modifications to the holographic  shear  sum rules in presence of 
 chemical potential has been obtained in \cite{David:2011hy}.
 This was done by considering charged blacks holes in gravitational 
 duals of the M2, D3 and M5-brane backgrounds. 
 The sum rules were modified due to  expectation values of operators
 in addition to the stress tensor due to the presence of chemical potentials
 Similar phenomenon was investigated for sum rules corresponding 
 to current correlators and other holographic models in 
 \cite{David:2012cd,WitczakKrempa:2012gn,Katz:2014rla,Myers:2016wsu}. 
 More recently the general structure of the shear sum rule for a  conformal 
 field theory in general dimensions was discussed in 
 \cite{Witczak-Krempa:2015pia}.

In this paper we derive the shear sum rule  for an arbitrary 
 conformal field theory in $d\geq 3$ dimensions using the general properties of 
 conformal field theories.
 We  consider the theory at finite temperature and at zero chemical potential. 
 We assume that the theory is such that at finite temperature,  it is only the 
 stress tensor that acquires a 
  non-zero expectation value. 
  Our main result for the shear sum rule is stated as 
\begin{eqnarray}\label{mainres}
 \lim_{\epsilon \rightarrow 0^+}\frac{1}{\pi}\int_{-\infty}^{\infty} \frac{\delta\rho(\omega)d\omega}{\omega -i\epsilon}&=&\left(\frac{(-1+d) d}{2 (1+d)}+\frac{(3-d) t_2}{2 (-1+d)}+\frac{\left(2+3 d-d^2\right) t_4}{(-1+d) (1+d)^2}\right)P, \nonumber\\
\rho(z) &=& \textrm{Im} G_R(\omega),\nonumber\\
G_R(t,x)&=& i\theta(t)[T_{xy}(t,x),T_{xy}(0)],\nonumber\\
\delta\rho(\omega)&=& \rho(\omega) -\rho(\omega)_{T=0}.
\end{eqnarray}
Here $P$ refers to the pressure of the theory and 
$t_2,t_4$  are the linearly independent parameters introduced 
 by Hofman and Maldacena  \cite{Hofman:2008ar}. 
 They carry the information of the three point function  of the 
 stress tensor and can be 
 written in terms of the constants $a, b, c$  of \cite{Osborn:1993cr} 
 \footnote{See equation (\ref{t2t4abc}) relating $t_2, t_2$ and $a, b, c$.}. 
 Note that pressure can be written in terms of the energy density $\epsilon = (d-1) P$. 
 
An immediate check on our sum rule is to apply it for theories which admit 
a two  derivative gravity dual.  From (\ref{mainres}) we see   that for these theories 
the sum rules reduces to 
\begin{eqnarray}
\lim_{\epsilon \rightarrow 0^+}\frac{1}{\pi}\int_{-\infty}^{\infty} \frac{\delta\rho(\omega)d\omega}
{\omega -i\epsilon}&=&\frac{d(d-1)P}{2(d+1)} .
\end{eqnarray}  
This is because $t_2 = t_4=0$ for these theories. 
We verify this  property by first explicitly using the 
data of the $3$ point functions  of stress tensors evaluated in \cite{Arutyunov:1999nw}. 
We then perform another check by evaluating the Greens function 
directly in the $AdS_{d+1}$ black hole following the methods developed 
in \cite{David:2011hy}. Thus a necessary but not sufficient condition 
for a theory to admit a $2$ derivative gravity dual is that the shear 
sum rule given in (\ref{mainres}) is satisfied. 

Positivity of energy flux \citep{Hofman:2008ar} or equivalently causality, constrains 
the parameters $t_2, t_4$.  These constraints were obtained in $d$ dimensions 
by \cite{Camanho:2009vw,Buchel:2009sk} and are stated in (\ref{t2t4abc}). 
Using these constraints we obtain  bounds on the shear sum rule 
for an arbitrary conformal field theory in $d$ dimensions. 
We also  apply the sum rules for specific theories in $d=3, 4,6$. 
We see that for  the M2-brane theory, ABJM theory, ${\cal N}=4$ Yang-Mills and the 
M5-brane theory, the coefficient involved in the sum rule is not renormalized as expected. 
For large $N$ Chern-Simons theory coupled to fundamental fermions  we 
evaluate the coefficient involved in the sum rule as a function of the t' Hooft coupling. 
We see that the bounds for the sum rule are saturated for the theory of free fermions
and free bosons.  As a simple  application of the sum rule on obtaining constraints 
on theories involving higher derivatives, we obtain the bounds on 
the coefficient of Gauss-Bonnet gravity in arbitrary dimensions from the sum rule.

Finally we study the high frequency behavior of the spectral density 
in the vector and the  scalar channels. That is we examine the spectral density 
corresponding to the  correlator  $\langle T_{tx} T_{xz} \rangle$  and  
$\langle T_{tt} T_{tt} \rangle $ respectively. 
We observe that the high frequency behavior is determined by 
Hofman-Maldacena coefficients corresponding to these channels. 

The organization of the paper is as follows. In section \ref{general} we present a 
general analysis of the shear sum rule we are interested in. 
Then we demonstrate  that the high frequency behavior of the 
shear correlator is determined by the Hofman-Maldacena coefficient
in the scalar channel and finally determine the sum rule. 
In section \ref{holographycheck} we perform a consistency check
on the sum rule using holography. We determine the sum rule using 
the evaluation of the three point function of the stress tensor 
in  $AdS_{d+1}$  and then compare it against the  direct evaluation of the sum rule 
from a black hole in $AdS_{d+1}$ and show agreement. 
In section \ref{bound} we re-write the sum rule in the Hofman-Maldacena variables
$t_2, t_4$ and obtain bounds on the sum rule  using causality. 
In section \ref{applications} we discuss applications of the sum rule
for well known theories in dimensions $d=3, 4, 6$.
In section  \ref{other} we study the high frequency behavior of 
the retarded Greens function in the vector and the sound channel and demonstrate that 
that it is determined by Hofman-Maldacena coefficients in the respective channels.
Section \ref{conclusion} contains the conclusions .
Appendix \ref{append} deals with the computation  of the Fourier transform 
of the OPE coefficients in the three channels. Appendix \ref{integrals} lists some 
integrals relevant for performing the Fourier transform. 
Finally Appendix \ref{sec:intro} reviews the evaluation of the three point function of the 
stress tensor in Chern-Simons vector models in the large $N$ limit. 

\section{The shear sum rule in conformal field theories}\label{general}

In this section we present the derivation of the sum rule in the shear channel for 
conformal field theories in dimensions $d> 2$. 
The derivation of spectral sum rules  rely on the analytical properties of the Greens function 
in the complex  $\omega$ plane. 
Consider a function $G(\omega)$ which is holomorphic in the upper half plane
including the real axis. 
For the present, let us assume that the function  has the following convergence 
property  in the upper half plane
\begin{equation} \label{conveg}
 \lim_{\omega \rightarrow  i \infty} G(\omega ) \sim \frac{1}{|\omega|^m}, \qquad m> 0 
\end{equation}
Then using Cauchy's theorem it can be shown that \footnote{See \cite{David:2011hy}) for example.}
\begin{equation}
 G(0) = \lim_{\epsilon \rightarrow 0^+} \int_{-\infty}^\infty \frac{d\omega}{\pi}
 \frac{\rho(\omega)}{ \omega - i \epsilon}, 
\end{equation}
where $\rho(\omega) = {\rm Im}\,  G(\omega)$. 
We will  restrict our attention to the retarded correlator of 
the $T_{xy}$ component of the stress tensor in  $d>2$  dimensions.
\begin{eqnarray}
G_R(t,x)=i\theta(t)\langle [T_{xy}(t,x),T_{xy}(0)] \rangle,
\end{eqnarray}
where the expectation value is taken in the theory held at finite temperature $T$. 
The Fourier transform is defined by 
\begin{eqnarray}
{G_R}(\omega, p) &=& \int d^dx e^{i\omega t - i p.\vec{x}} G_R(t,x).
\end{eqnarray}
We will be interested  in the sum rule for the spectral function at $p=0$,  defined as  
\begin{eqnarray}
\rho(\omega ) &=& \textrm{Im}{G}_R(\omega, 0).
\end{eqnarray}

Let us now examine if each of the assumptions involved in deriving the sum rule is satisfied 
for the  the retarded correlator.  
The  physical reason why the retarded correlator is analytic in the upper half plane is 
causality. It is easy to see this  from the inverse Fourier transform
\begin{equation}
 G_R(t ) = \int \frac{d\omega}{2\pi} e^{ - i \omega t} G_R(\omega). 
\end{equation}
For $t<0$ and only when $G_R(\omega)$ is holomorphic, the contour can be closed in the 
upper half plane resulting in $G_R(t<0)=0$ which is a requirement for the 
retarded correlator.
Now for conformal field theories the property  (\ref{conveg}) is not satisfied. 
As we will see below the retarded correlator 
diverges as $\lim_{\omega\rightarrow \infty} G_R(\omega) \sim \omega^d$. 
However we can still define a regularized Greens function $\delta G_{R}$ which 
satisfies the property (\ref{conveg}) for which we can 
apply Cauchy's theorem and obtain
\begin{equation}
 \delta G_R(0) = \lim_{\epsilon \rightarrow 0^+} \int \frac{d\omega}{\pi}
 \frac{\delta \rho(\omega)}{ \omega - i \epsilon}, 
\end{equation}
where $ \delta \rho(\omega) = {\rm Im }  ( \delta G_R(\omega) ) $.
The precise definition of the regularization of course depends on the 
details of the high frequency behaviour.

Thus the thing we need to do for the  derivation of  the sum rule is to examine the 
high frequency behavior of the shear correlator in the upper half plane. 
For this we  continue $G_R(\omega)$ into the upper half plane using the 
following relation between the retarded correlator and the 
Euclidean correlator which can be proved from the definition of these correlators
\footnote{See  for example \cite{Meyer:2011gj} for a proof.}
\begin{equation}
 G_R( i 2\pi n T) = G_E(2\pi n T). 
\end{equation}
Here $G_E$ is the Euclidean time ordered correlator and $2 \pi n T$ is the Matsubara
frequency. 
This relation provides a distinguished analytic continuation 
\begin{equation}\label{analyt}
 G_R(i\omega) = G_E(\omega).
\end{equation}
We need the behavior of the retarded Greens function as $\omega\rightarrow\infty$. 
Consider the Euclidean correlator in position space. For time 
intervals $\delta t \ll \beta = \frac{1}{T}$,
the operator product expansion (OPE) of the stress tensor offer a good 
asymptotic expansion. Therefore for $\omega \gg T$, we can replace the 
Euclidean correlator by its OPE. This allows us to obtain the asymptotic 
behaviour of the $G_R(i\omega) $ as $\omega\rightarrow \infty$. 
To conclude,  the strategy is to write down the leading terms of the 
Euclidean OPE of the stress tensor and then Fourier transform each of these
terms to frequency space to obtain the large $\omega$ behavior. 
Such an analysis has been used earlier in \cite{Romatschke:2009ng,CaronHuot:2009ns,Witczak-Krempa:2015pia}.

Let us first use  the OPE of the stress tensor in the Euclidean two point function. 
Using the result of \cite{Osborn:1993cr} we obtain 
\begin{eqnarray}\label{OPE}
\langle T_{xy}(s)T_{xy}(0)\rangle \sim C_T\frac{I_{xy,xy}(s)}{s^{2d}} + 
\hat{A}_{xyxy\alpha\beta}(s)  \langle T_{\alpha\beta}(0) \rangle + \cdots
\end{eqnarray}   
Here the OPE has been sandwiched between thermal states. 
$s$ refers to the position in $d$ dimensions and $C_T$ is a constant which determines the 
normalization of the $2$ point function of the stress tensor.
By dimensional analysis, 
the tensor structure $I_{\mu\nu\rho\sigma}$  is dimensionless, 
while the $\hat A_{\mu\nu\rho\sigma}$ scales like $1/s^d$.
From this scaling property it is easy to conclude that  the following integral 
scales as 
\begin{eqnarray}\label{defI}
 \int  d^d x e^{i \omega t } C_T \frac{ I_{xy, xy} (x) }{ |x^{2d}| }\equiv {\cal I}
 \sim \omega^d 
 \log( \frac{\omega}{\Lambda} ) ,  
\end{eqnarray}
where $\Lambda$ is a cut off in the integration 
\footnote{We choose this  branch cut to lie in the lower half  $\omega$ plane which 
ensures that the correlator is holomorphic in the upper half plane.}. 
As we will soon see, we will not need the details of this divergent term. 
Lets examine the second term
\begin{eqnarray}\label{defJ}
\int_{-\infty}^\infty d^d x e^{i\omega t } \hat A_{xyxy\alpha\beta}(x) \langle
T_{\alpha\beta} \rangle \equiv {\cal J}  \sim 
\omega^0 \hat a^{\alpha\beta} \langle T_{\alpha\beta} \rangle, 
\end{eqnarray}
 where $\hat a^{\alpha\beta}$ are $O(1)$ coefficients. 
 For the thermal vacuum, the one point functions  $\langle T_{\alpha\beta} \rangle$  can be written in terms of 
 the  energy density and the pressure.  Thus these terms also contributes in the $\omega \rightarrow \infty$ limit
 We assume that there are no other operators of conformal dimensions $\Delta \leq d$ which gain 
 expectation value in the thermal vacuum.   The rest of the terms in the OPE  (\ref{OPE})  involve
 operators of dimensions $\Delta > d$ whose terms are suppressed in the $\omega \rightarrow \infty$
 limit as $O(1/\omega^{\Delta -d} )$. 

 Thus we need to regularize the retarded Greens function by 
subtracting ${\cal I}$ and ${\cal J}$ defined in (\ref{defI}) and (\ref{defJ}). 
From examining  ${\cal I}$, we see that it is identical to the Fourier transform 
of the retarded  Greens function at zero temperature \footnote{Note that 
though the OPE 
is in Euclidean space,  we are examining the limit  $\omega\rightarrow i\infty$
of retarded greens function using the relation (\ref{analyt}).}. 
Therefore we define the regularized Greens function as 
\begin{equation}
 \delta G_R(\omega) = G_R(\omega)|_T - G_R(\omega)|_{T=0} - {\cal J}. 
\end{equation}
Now this definition ensures that when $\omega \rightarrow i\infty$, the regularized Greens function 
behaves as 
\begin{equation}
 \lim_{\omega\rightarrow i\infty }\delta G_R(\omega)  \sim O( \frac{1}{\omega^{\Delta -d} } ) ,
\end{equation}
and we can apply Cauchy's theorem to obtain the sum rule
\begin{eqnarray}
 \lim_{\epsilon \rightarrow 0^+} \int_{-\infty}^\infty \frac{d\omega}{\pi}
 \frac{\delta \rho(\omega)}{ \omega - i \epsilon}  &=&  \delta G_R(0) , \\ \nonumber
 &=&  G_R(0)|_T - G_R(0)|_{T=0} - {\cal J}. 
\end{eqnarray}
Now let us assume hydrodynamic behaviour of the theory at small wavelengths. 
This implies that we can identify the zero frequency behaviour of the retarded
Greens function with pressure \footnote{We define $\langle T_{xy} \rangle  = - \frac{1}{\sqrt{g}}\frac{\partial \ln Z}{\partial g_{xy} }$, this along with the constitutive 
relation for the stress tensor from hydrodynamics leads to $ G_R(0)|_T = P$.
Note that  in general, 
it is important to obtain the zero frequency behaviour from hydrodynamics.  
Naive use of the analytical continuation from Euclidean 
Greens function at zero frequency could miss delta function contributions
at which occur at non-zero values of chemical potentials. We thank
Zohar Komargodski for raising this issue.} 
\begin{equation} \label{hydro}
 G_R(0)|_T = P. 
\end{equation}
Also we have the property 
\begin{equation} \label{zero}
 G_R(0)|_{T=0} = 0 ,
\end{equation}
which arises from fact that zero temperature retarded Greens function 
vanish, since the pressure vanishes at zero temperature in a conformal field theory. 
Combining (\ref{hydro}) and (\ref{zero}) we obtain the sum rule
\begin{eqnarray}\label{sumrcom}
 \lim_{\epsilon \rightarrow 0^+} \int_{-\infty}^\infty \frac{d\omega}{\pi}
 \frac{\delta \rho(\omega)}{ \omega - i \epsilon} = P - {\cal J},
\end{eqnarray}
where
\begin{eqnarray}
\delta \rho(\omega) &= &{\rm Im} G_R(\omega)|_T - {\rm Im} G_R(\omega)|_{T=0} ,\\ \nonumber
&=& \rho(\omega)_{T}  - \rho(\omega)|_{T=0}.
\end{eqnarray}
This is because ${\cal J}$ will turn out to be real and will not contribute to the
spectral density. 
It is important to note the  origin of the $2$ terms in the RHS of the sum rule (\ref{sumrcom}). 
The first term is due to the long wave length hydrodynamic behavior of the 
theory, while the 2nd term is due to the short distance behavior of the theory 
and results from the OPE. 

\subsection{High frequency behavior and Hofman-Maldacena coefficient}

In this section we will evaluate ${\cal J}$ for an arbitrary conformal field theory 
in $d$ dimensions. Lets recall its definition 
\begin{equation}
{\cal J} (\omega, p=p_z) = \int d^d x e^{i\omega t  - i  p  z} 
 \hat A_{xyxyxy\alpha\beta} (x) \langle T_{\alpha\beta} \rangle,
\end{equation}
where we have also introduced momentum $p=p_z $ along 
the the direction orthogonal to $x, y$ in the Fourier transform. 
In the end we will set $p=0$, or equivalently expand the ${\cal J}$ at 
$\omega \rightarrow \infty$ and extract the constant term. 
The only components of the  expectation value of the stress tensor
which are non-zero are given by 
\begin{equation}
\langle T_{tt} \rangle= \epsilon_E, \qquad  
\langle T_{ij} \rangle = P\delta_{ij} , \quad i, j  = 2 \cdots d.
\end{equation}
The subscript in $\epsilon_E$ reminds us that that this is the energy density in the Euclidean
theory We have  $\epsilon_E = - \epsilon$ where $\epsilon$ is the energy density in 
the Minkowski theory.  Also recall that from  conformal invariance we have the relation
\begin{equation}
\epsilon = (d-1) P.
\end{equation}
Finally the tensor structure $\hat A_{\mu\nu\rho\sigma} (s) $ is given by 
\cite{Osborn:1993cr}
\begin{eqnarray}\label{tenstu}
\hat{A}_{\mu\nu\rho\sigma\alpha\beta} C_T &=& \frac{(d-2)}{d+2}(4a+2b-c)H^1_{\alpha \beta \mu\nu\rho\sigma}(s) +  \frac{1}{d}(da+b-c) H^2_{\alpha\beta \mu\nu\rho\sigma}(s) \nonumber\\
&&-\frac{d(d-2)a-(d-2)b-2c}{d(d+2)} (H^2_{\mu\nu\rho\sigma\alpha\beta}(s)+H^2_{\rho\sigma\mu\nu\alpha\beta}(s))+\frac{2da+2b-c}{d(d-2)} H^3_{\alpha\beta \mu\nu\rho\sigma}(s)\nonumber\\
&&-\frac{2(d-2)a-b-c}{d(d-2)}H^4_{\alpha\beta \mu\nu\rho\sigma}(s)-\frac{2((d-2)a-c)}{d(d-2)}(2H^3_{\mu\nu\rho\sigma\alpha\beta}(s)\nonumber\\
&&+\frac{((d-2)(2a+b)-dc)}{d(d^2-4)}(H^4_{\mu\nu\rho\sigma\alpha\beta}+H^4_{\rho\sigma\mu\nu\alpha\beta})(s))\nonumber\\
&&+(C h^5_{\mu\nu\rho\sigma\alpha\beta}+D(\delta_{\mu\nu} h^3_{\rho\sigma \alpha \beta}+ \delta_{\rho \sigma} h^3_{\mu \nu \alpha \beta}))S_d \delta^d(s),\nonumber\\
&=& I_1+I_2+I_3+I_4+I_5+I_6+I_7+I_8,
\end{eqnarray}
where $C_T, C $ are functions of the parameters $a, b, c$ which determine the 
three point function of the stress tensor  in the conformal field theory  and 
are given by 
\begin{eqnarray}
&&C_T = \frac{8\pi^\frac{d}{2}}{\Gamma(\frac{d}{2})} \frac{(d-2)(d+3)a-2b-(d+1)c}{d(d+2)}, 
\\ \nonumber
&&C =\frac{(d-2)(2a+b)-dc}{d(d+2)} ,  \qquad 
S_d = \frac{ 2\pi^{\frac{d}{2}} }{\Gamma( \frac{d}{2})}.
\end{eqnarray}
The detailed evaluation of the Fourier transform to obtain ${\cal J}$
 is tedious and is  provided in 
appendix  \ref{fouriertransformtensor}. For some intuition we will present the Fourier transform 
of the term $I_4$, note that we only need the diagonal entries $\alpha=\beta$. 
\begin{eqnarray}
I_4(s)&=&\frac{2da+2b-c}{d(d-2)} H^3_{\alpha\beta xyxy}(s) , \quad
H^3_{\alpha\beta xyxy}(s)=(\partial_\alpha \partial_\beta-\frac{1}{d}\delta_{\alpha\beta} \partial^2) \frac{1}{(t^2+\vec r^2)^\frac{d-2}{2}}, \nonumber \\ 
I_4(\omega, p)&=&\frac{2da+2b-c}{d(d-2)} H^3_{\alpha\beta xyxy}(\omega, p).
\end{eqnarray}
Performing the Fourier transform we obtain 
\begin{eqnarray}
\sum_{i=1}^{d-1} H^3_{ii xyxy}(\omega, p)
=\frac{4 \pi ^{d/2} \left(d \omega ^2-p^2-\omega ^2\right)}{d \Gamma 
\left(\frac{d}{2}-1\right) \left(p^2+\omega ^2\right)} = - 
H^3_{tt xyxy}(\omega, p). 
\end{eqnarray}
Including the coefficients  obtained from the  expectation value of the 
stress tensor we obtain 
\begin{eqnarray}
\hat I_4(\omega, p)=\frac{(2 a d+2 b-c)}{d (d-2)} 
 \left(\frac{ 4 \pi ^{d/2} \left(d \omega ^2-p^2-\omega ^2\right)}{d \Gamma \left(\frac{d}{2}-1\right) \left(p^2+\omega ^2\right)}
 \right) (P-\epsilon_E) .
\end{eqnarray}
At the end we need to take $\hat I_4(\omega,  0)$ as the contribution to ${\cal J}(\omega, 0)$. 
Similarly we need to perform the Fourier transform for each of the terms
$I_1, \cdots I_7$ in (\ref{tenstu}). 
Let us briefly indicate the procedure involved in performing the Fourier transform.
We first parametrize the  spatial directions   in terms of polar co-ordinates and 
then perform the angular integrations. We then perform the radial integration and 
finally the integration over time $t$. 
In each of the integrals involved,  we have verified that that interchange of the 
order of the time and the radial integrations does not affect the results. 
The details of each of the Fourier transform is given in the appendix \ref{fouriertransformtensor}. 
Thus sum of all these terms are given by 
\begin{eqnarray}\label{malcoeff}
\sum_{l=1}^7 \hat I_l (\omega, 0) = 
\frac{(1-d)  (a (d (d+4)-4)+d (2 b-c))}{2 \left(-a \left(d^2+d-6\right)+2 b+c d+c\right)} P.
\end{eqnarray}
Here we have used  $\epsilon_E = -(d-1) P$ to write all terms in terms of
the pressure. 
At this point we observe that  the term in (\ref{malcoeff}) is proportional to the
Hofman-Maldacena coefficient  obtained by 
examining positivity of energy flux in the spin zero channel \cite{Hofman:2008ar}
\footnote{We refer to the combination of the stress tensor OPE coefficients 
which occur in specifically in various channels `Hofman-Maldacena coefficients', 
though the OPE's themselves where known earlier by \cite{Osborn:1993cr}.}. 
More precisely 
for arbitrary dimensions $d$ we obtain the relation 
\begin{equation}\label{malcoef}
\sum_{l=1}^7 \hat I_l (\omega, 0)  = {2(1-d)} P   a_{T, 0},
\end{equation}
where $a_{T, 0}$ is defined in equation (2.16) of \cite{Komargodski:2016gci} and is given 
by 
\begin{equation}\label{malcoef0}
a_{T,0} = \frac{1}{4} \frac{  a (d (d+4)-4)+d (2 b-c)}{  -a \left(d^2+d-6\right)+2 b+c d+c }.
\end{equation}
Note  that  \cite{Komargodski:2016gci} 
\footnote{See \cite{Kulaxizi:2010jt} for 
earlier work in this direction.}
 were examining the  kinematic regime
of space like momenta, while we are examining the situation at vanishing spatial momenta but 
non-zero frequency which is relevant for the sum rule. 
Here we are examining the expectation value of the stress tensor in a thermal state while 
 \cite{Komargodski:2016gci} looked at the expectation value in a single particle. 
It is interesting that for both these situations  we obtain  the Hofman-Maldacena coefficient. 
Finally note  $\ref{tenstu}$ is a contact term in position space responsible
for ensuring the conformal Ward identity \cite{Osborn:1993cr}. 
Its Fourier transform is given by 
\begin{equation}
\hat I_8 (\omega, p) = \frac{((2 a+b) (-2+d)-c d) P}{-2 b-c (1+d)+a \left(-6+d+d^2\right)}.
\end{equation}
Summing all the individual contributions in ${\cal J}$ we obtain 
\begin{eqnarray}\nonumber
{\cal J}  &=& \frac{(1-d) P (a (d (d+4)-4)+d (2 b-c))}{2 \left(-a \left(d^2+d-6\right)+2 b+c d+c\right)}
 + \frac{((2 a+b) (-2+d)-c d) P}{-2 b-c (1+d)+a \left(-6+d+d^2\right)}. \\ 
\end{eqnarray}
We are now in a position to evaluate $P -{\cal J}$ and write down the 
sum rule of conformal field theories in arbitrary dimensions
\begin{eqnarray} \label{sumabc} \nonumber
&& \lim_{\epsilon \rightarrow 0^+} \int_{-\infty}^\infty \frac{d\omega}{\pi}
 \frac{\delta \rho(\omega)}{ \omega - i \epsilon} = \delta G(0) , \\
&=&
\frac{\left(2 c+d (c+2 b d-c d)+a \left(8+d \left(-6+d+d^2\right)\right)\right) P}{2 (2 b+c+c d)-2 a \left(-6+d+d^2\right)} .
\end{eqnarray}

Let us now discuss the important assumption used in arriving at the above sum rule (\ref{sumabc}). 
As mentioned around equation (\ref{defJ}) we have assumed that no other operator of 
dimensions $\Delta \leq d$ gains expectation value in the thermal vaccum. 
This assumption   holds  true for theories which admit a pure $AdS_{d+1}$ dual, for example 
${\cal N}=4$ Yang-Mills. However if one turns on chemical potential for 
$R$-charges in  such theories, other  marginal operators 
are turned and and the sum rule is corrected. 
Such corrections modify the RHS of the sum rule and 
have been derived holographically in \cite{David:2011hy} for the shear sum rule 
and \cite{David:2012cd} for sum rules obeyed by current correlators. 
But, there are also situations in which relevant operators can be turned on, for example
for the quantum Ising model in $2+1$ dimensions \footnote{We thank 
William Witczak-Krempa for bringing this model to our attention and 
for correspondence which brought out these  points.}. In such situations 
one needs to subtract the expectation value of the 
corresponding operator multiplied by its OPE coefficient in the integrand 
which occurs in the LHS of the sum rule \cite{Witczak-Krempa:2015pia} 
\footnote{See equation (12) of  \cite{Witczak-Krempa:2015pia}.}. 
Such a situation was also encountered holographically in \cite{David:2012cd} for the case of 
M2 branes at finite  R-charge. 
Thus these situations will necessitate the inclusion of   additional terms in the sum rule. 
However the term in the RHS found in (\ref{sumabc}) will not be modified. 
We would also like to emphasise though the general structure of the 
term in the RHS can be argued by scaling and the OPE, we have related the 
constant in front the pressure term to the data of the conformal field theory.  

Let us now briefly discuss the possible modifications of the sum rule 
for non conformal field theories of which QCD is a relevant example. 
In the derivation of the sum rule, there are two terms that 
contribute to the RHS  ${\cal J}$ and $G_R(0)|_T =P$ in (\ref{sumrcom}). 
The coefficient of  the expectation value  of the stress energy tensor in 
${\cal J}$  just depends on the OPE coefficients of the theory at high 
energy. Therefore in a theory like QCD which is asymptotically free and
conformal at high energies, these coefficients can be evaluated perturbatively. 
However the infrared term  $G_R(0)|_T =P$ is much more difficult to 
obtain and rely on  the hydrodynamic behaviour of the theory. 
Further more in all our manipulations especially in evaluating the 
one point function of the stress energy tensor we have used 
the relation $\epsilon = (d-1) P$ which is true only in conformal field theories. 
Also note that these expectation values depend on coupling and are 
evaluated at low energies of the theory.
Thus taking these inputs in one can evaluate sum rules for non-conformal field theories. 
QCD is an example where such sum rules have been evaluated \cite{Romatschke:2009ng}. 
Here one can see in the analysis
\footnote{See equation (37) of \cite{Romatschke:2009ng}.}, the coefficient ${\cal J}$ is 
evaluated perturbatively. However, there are other possible  marginal operators 
that can gain expectation values in the thermal vacuum eg. ${\rm Tr } F^2$. 
These in principle depend on $\epsilon - 3P$ and have not yet been carefully 
evaluated \footnote{The sum rule for the bulk channel in QCD has also been 
evaluated in \cite{Romatschke:2009ng}. Here too the high energy contribution
can be determined from a perturbative calculation.}.
Therefore in conclusion, the term involving the high energy 
there point function ${\cal J}$ in the sum rule  can be obtained using a perturbative analysis,
however there are usually other  terms that are sensitive to the non-conformal 
nature of the theory and one needs to analyse them carefully.

\section{Check from  holography}\label{holographycheck}

In this section we evaluate the $\delta G_R(0)$, the  RHS of the sum rule in (\ref{sumabc}) 
in holography.  We do this in two  different ways. 
First we use the values of the parameters $a, b, c$ obtained by evaluating the 
three point function of stress tensor from $AdS_{d+1}$ 
 in \cite{Arutyunov:1999nw} and determine
$\delta G_R(0)$. We then evaluate $\delta G_R(0)$ more directly by considering a black hole
in $AdS_{d+1}$ and evaluating the retarded Green's function. 
In both cases the answer reduces to 
\begin{equation}
\delta G_R(0) =  \frac{d}{ 2 ( d+1) } \epsilon.
\end{equation}
This provides a consistency check within holography of the sum rule in 
(\ref{sumabc}).

\subsection{ $\delta G_R(0)$ using the 3-point function  of stress tensor}

The parameters $a, b, c$ which determine   the three point function of stress tensor
evaluated in $AdS_{d+1}$ are usually given in terms of ${\cal A}, {\cal B}, {\cal C}$ which 
are linearly related to $a, b, c$ \cite{Arutyunov:1999nw, Bastianelli:1999ab}

\begin{equation} \label{calabcd}
a=\frac{\mathcal{A}}{8}, \qquad 
b = \frac{\mathcal{B}-2\mathcal{A}}{8}, \qquad
c = \frac{\mathcal{C}}{2}.
\end{equation}                                 
The  three parameters $\mathcal{A}, \mathcal{B}$ and $\mathcal{C}$ in 
general $AdS_{d+1}$ upto the overall Newton's constant in $d+1$ dimensions 
are  given by \citep{Arutyunov:1999nw}
\footnote{See equations (3.24), (3.26) of  \cite{Arutyunov:1999nw}. } 
\begin{eqnarray}\label{defcalabc}
\mathcal{A} &=& 3 \Delta_d a_1, \qquad \mathcal{B} = \Delta_d (2a_1+a_2-2a_3) \nonumber\\
\mathcal{C} &=& \Delta_d (2 a_4 + a_5).           
\end{eqnarray}                        
where,
\begin{eqnarray}
\Delta_d  &=& \frac{d \Gamma\left(d\right)}{2 \pi^d (d-1)^2}, \qquad 
a_1 = \frac{4 d^3}{3-3 d},\nonumber\\
a_2 &=& -\frac{4 \left(6+d \left(-4-6 d+3 d^3\right)\right)}{3 (-1+d)}, \qquad 
a_3 = -\frac{2 (6+d (-7+d (-6+5 d)))}{3 (-1+d)} ,\nonumber\\
a_4 &=& \frac{1}{3} \left(-1+\frac{1}{-1+d}+4 d-3 d^3\right), \qquad 
a_5 = \frac{1}{3} \left(5+\frac{1}{-1+d}-5 d\right) .\nonumber\\
\end{eqnarray} 
Now using (\ref{defcalabc}) we find the the parameters determining the 
three point function of the stress tensor are given by 
\begin{eqnarray}
a &=&  -\frac{d^4 \pi ^{-d} \Gamma[d]}{4 (-1+d)^3}, \nonumber\\
b &=& -\frac{d \left(1+(-3+d) d^2\right) \pi ^{-d} \Gamma(1+d)}{4 (-1+d)^3},\nonumber\\
c &=& \frac{d^3 (1-2 (-1+d) d) \pi ^{-d} \Gamma(d) }{4 (-1+d)^3}.\nonumber\\
\end{eqnarray}
Finally substituting these values into the coefficient $\delta G_R(0)$  of the sum rule 
given in (\ref{sumabc}) we obtain 
\begin{eqnarray}\label{minimalscalar}
\delta G_R(0)  &=& \frac{\left(2 c+d (c+2 b d-c d)
+a \left(8+d \left(-6+d+d^2\right)\right)\right) P}{2 (2 b+c+c d)-2 a \left(-6+d+d^2\right)},\nonumber\\
&=& \frac{(-1+d) d P}{2 (1+d)}, \nonumber\\
&=& \frac{d \epsilon}{2(d+1)}.
\end{eqnarray}
Thus the relatively complicated rational expression in terms of $a, b, c$ reduces to a simple expression 
which just depends on the dimension $d$. 

\subsection{ $\delta G_R(0)$ from the  $AdS_{d+1}$  black hole}

In this section we evaluate   $\delta G_R(0)$  directly by  studying  the retarded Green's 
function in the $AdS_{d+1}$ black hole background. 
This method was originally used for obtaining the holographic 
sum rule in $d=4$ dimensions in \cite{Romatschke:2009ng} and then further in 
\cite{Gulotta:2010cu,David:2011hy,David:2012cd} for other dimensions and more general situations. 
We will follow the systematic approach developed in  \cite{David:2011hy}.
The geometry dual to a conformal field theory  in $d$ dimensions at finite temperature $T$ is given by the 
Schwarzschild black hole in $AdS_{d+1}$  at the Hawking  temperature $T$.  
The metric  and the Hawking temperature is  given by \footnote{See for example 
in \cite{Bhattacharyya:2008mz}.}. 
\begin{eqnarray}\label{AdSd+1}
ds^2 &=& \frac{dr^2}{r^2 f(r)} + r^2(-f(r)^2dt^2+dx_1^2\cdots + dx_{d-1}^2 ),\\ \nonumber
f(r) &=& 1-\left(\frac{r^+}{r}\right)^d, \qquad 
T = \frac{d r_+}{4\pi r}.
\end{eqnarray} 
$r_+$ is the radius of the Horizon and 
for convenience we have set the radius of $AdS_{d+1}$ to unity. 
The retarded shear correlator is obtained by consider the fluctuation $\delta g_{xy}$ of the metric which 
is dual to the stress tensor $T_{xy}$. 
Let us define  the fluctuation as 
\begin{equation}
 \delta g_{xy} = r^2\phi(r)e^{i\omega \tau}.
\end{equation}
This  perturbation  obeys the 
equation of motion of a massless scalar in the background, which is given by 
\begin{eqnarray}
&& \frac{d^2 \phi}{dr^2}  + (\frac{d-1}{r}+\frac{F'}{F})\frac{d \phi }{dr}+ \frac{\omega^2}{F^2} \phi =0, \\
&& F=r^2f(r).
\end{eqnarray}
The retarded Greens function is obtained by imposing in going boundary conditions 
at the horizon $r_+$ and obtaining the solution to $\phi$ at the boundary $r \rightarrow \infty$ \cite{Son:2002sd}.
The  Greens function is given by 
\begin{eqnarray}\label{gravretcorr}
G_{xy,xy}^R (\omega, T)&=& \hat G(\omega, T)   + G_{\rm{counter}} (\omega, T)
+ G_{\rm{contact} } ( T) ,
\nonumber\\
 \hat G(\omega, T) &=& \frac{-1}{2\kappa^2} \lim_{r\rightarrow\infty}
 \frac{F r^{d-1} \phi'}{\phi} ,
\end{eqnarray}
where  $\kappa$ is the gravitational coupling constant.  $G _{ \rm{counter}} (\omega )$ are the counter 
terms which are necessary to remove the $\log(r)$ divergences. 
These are independent of the 
temperature  and are identical to the counter terms one needs at $T=0$. Hence 
they cancel on considering  $\delta G_R(\omega)$.  
Now the contact term in the Greens function arises from the frequency independent 
term in the effective action and is given by 
\begin{equation}
 G_{\rm{contact} } ( T)   = P .
\end{equation}
Thus on considering $\delta G_R(0 )$, both the counter term as well as the contact 
term cancel.  The counter terms cancel since they are temperature independent, while 
the contact term cancels because at zero frequency we have  $\hat G_R (0, T) = P$. 
The terms which contribute to $\delta G_R(0 )$   arise from the temperature dependent 
terms which are  finite in the high frequency $\omega \rightarrow i \infty$ limit   of $\hat G_R (0, T)$. 
This leads us to  study the function 
 $\hat G(\omega, T)$ as $\omega \rightarrow \infty$. It is convenient to introduce the variables
\begin{eqnarray}\label{chgvar}
i\lambda = \frac{\omega}{r_+}, \qquad y= \frac{ \lambda r_+}{r} .
\end{eqnarray}
In these variables we are examining the limit $\lambda\rightarrow \infty$. 
 The equation for $\phi$ becomes,
\begin{eqnarray}\label{diffeq1}
\frac{d^2\phi}{dy^2}  - \frac{1}{f(y)y}\left(d-1+\frac{y^d}{\lambda^d}\right) 
\frac{d \phi}{dy}  -\frac{\phi}{f^2(y)}=0,
\end{eqnarray}
where,
\begin{eqnarray}
f(y)=1-\frac{y^d}{\lambda ^d}.
\end{eqnarray}
Expanding the differential equation (\ref{diffeq1} as a   a power series in $\lambda$ we obtain  
\begin{eqnarray}\label{diffeq1exp}
\frac{d^2 \phi}{dy^2}  - (\frac{d-1}{y}+ \frac{dy^{d-1}}{\lambda^d} 
+ \cdots) \frac{d \phi}{dy} -(1+\frac{2 y^d}{\lambda^d}+\cdots)\phi=0.
\end{eqnarray}
It's easy to see that the leading equation in $\lambda$ is the equation of 
the minimally coupled scalar in pure $AdS_{d+1}$, the sub-leading terms in $\lambda$ account for the
presence of the black hole. 
Lets us solve for the function $g_R(\omega)$
 perturbatively in $\frac{1}{\lambda}$. We define 
\begin{eqnarray} \label{gexpand}
g = \frac{\phi'(y)}{\phi(y)} 
= \sum_{n=0}^\infty \frac{g_n}{\lambda^{nd}}.
\end{eqnarray}
Substituting the expansion for $g$ in
 \eqref{diffeq1exp} and matching  terms order by order in $\frac{1}{\lambda}$ we obtain the 
 following equations for the leading terms. 
\begin{eqnarray}
&&g_0' -g_0^2-\frac{d-1}{y} g_0 -1 = 0,\nonumber\\
&&g_1'+(2g_0 -\frac{d-1}{y}) g_1 -dg_0y^{d-1} -2y^d = 0 .
\end{eqnarray}
Note that that equation for the zeroth order term $g_0$ is identical to that 
obtained in the pure $AdS_{d+1}$ background. 
The two solutions to $g_0$, given by,
\begin{eqnarray}\label{gravitysoln}
g_0^{(1)} &=& -\frac{K_{\frac{d-2}{2}}(y)}{K_{\frac{d}{2}}(y)}, \qquad 
g_0^{(2)} = -\frac{I_{\frac{d-2}{2}}(y)}{I_{\frac{d}{2}}(y)}.
\end{eqnarray}
Now the second solution corresponds to the 
solution of  $\phi$ which diverges at the origin  $y \rightarrow \infty$.  Therefore in the 
strict $\lambda \rightarrow \infty$ we must discard the second solution. 
In \cite{David:2011hy}, it has been shown in detail that the second solution does 
not contribute to the finite term as $\lambda\rightarrow \infty$. 
Thus we look for the expansion of the first solution, substituting this 
solution into the 
 the second equation we obtain $g_1^{(1)} $
\begin{eqnarray}
g^{(1)}_1 &=& \frac{1}{yK_{\frac{d}{2}}(y)^2} \int_0^y (2y^d+g_0 d y^{d-1})yK_{\frac{d}{2}}(y) + \frac{C}{yK_{\frac{d}{2}}(y)^2} ,
\end{eqnarray}
where the term containing the arbitrary constant $C$ is the homogeneous solution. 
We set this 
constant to zero since the presence of $C$ makes $g^{(1)}_1$ grow exponentially 
at $y\rightarrow\infty$ and thus alters the boundary condition set by $g^{(0)}_1$. 
Therefore we have 
\begin{eqnarray}
g^{(1)}_1 &=& \frac{1}{yK_{\frac{d}{2}}(y)^2} \int_0^y (- d y^d K_{\frac{d}{2}}(y)K_{\frac{d-2}{2}}(y)+2y^{d+1}K_{\frac{d}{2}}(y)^2 ),\nonumber\\
&=& \frac{d y^{d-1}}{2} + \frac{y^{d+1}}{d+1} 
\left(1 - \frac{K_{\frac{d+2}{2}}(y)^2}{K_{\frac{d}{2}}(y)^2}\right).
\end{eqnarray}  
Here  we have integrated each of the terms in the integrand by parts and then 
we  use relations for the derivatives of the Bessel functions. 
Now to obtain the retarded Greens function we need the behavior of 
$g^{(1)}$ close to the boundary $y\rightarrow 0$, which is given by 
\begin{eqnarray}
\lim_{y\rightarrow 0} g^{(1)}_1(y)  = 
-\frac{d (d-1) }{2 (1+d) }y^{d-1} +\frac{(2+d) }{2+d-d^2} y^{d+1}  + O(y^{d+3}).
\end{eqnarray}
Now substituting  the expansion  (\ref{gexpand})  into the definition of the 
Greens function we obtain
\begin{equation}\label{greenhigh}
\lim_{\lambda \rightarrow \infty} \hat G (\omega, T) = \frac{r_+^d}{2\kappa^2} 
 \lim_{y \rightarrow 0} 
\left( \frac{ \lambda^d}{ y^{d-1} } g_0^{(1)} - \frac{ d( d-1) }{ 2 ( d+1) } \right)  + 
O(\frac{1}{\lambda^d}),
\end{equation}
where we have used the change of variables in (\ref{chgvar}) from $r$ to $y$. 
Note that the leading term is proportional to $(\lambda r_+)^d = (-i\omega)^d$ 
and therefore independent of temperature. 
This term will be present also in pure $AdS_{d+1}$ and therefore 
will cancel on considering $\delta G_R(\omega)$. 
Thus the  finite term  in the high frequency limit 
in addition to the contact term that  one needs to subtract to regularize 
the Greens function is the second term in (\ref{greenhigh}). 
As we have argued before the  the contact term is independent of the frequency 
and will cancel since $G_R(0) = P$.  Using all these inputs  we have
\begin{equation}
\delta G_R(0) =   \frac{r_+^d}{2\kappa^2} \frac{d(d-1)}{2(d+1)} .
\end{equation}
We can now rewrite this expression in terms for the 
field theory variables using  the relation for pressure  \cite{Bhattacharyya:2008mz}
\begin{equation}
P = \frac{r_+^d}{2\kappa^2}.
\end{equation}
This results in 
\begin{equation} \label{holsum}
\delta G_R(0) =  \frac{d(d-1)}{2(d+1)}  P =  \frac{d }{2(d+1)}  \epsilon.
\end{equation}
The above result for the holographic shear sum rule was first derived in \cite{Gulotta:2010cu}. 
The expression for $\delta G_R(0)$ in (\ref{holsum})  coincides 
 with the evaluation of $\delta G_R(0)$ in the 
previous section given in (\ref{minimalscalar}), 
which used the input from the three point function 
of the stress tensor evaluated holographically. 
However the evaluation of $\delta G_R(0)$ in this section 
did not rely on the explicit expression given in 
 given in (\ref{sumabc})  in terms of the parameters of the three point function $a, b, c$
 and therefore provides a consistency check
 for the general  sum rule derived in (\ref{sumabc}).

 \section{Sum rule, Hofman-Maldacena variables, causality bounds}
 \label{bound}
 
 The sum rule in (\ref{sumabc})  is a  ratio of a linear combination of the 
 constants $a, b, c$ which determine the  three point function. 
 The variables $t_2, t_4$ which were used to characterize the 
 positive of energy flux \cite{Hofman:2008ar} were also related to the ratios of 
 linear combinations of the constants $a, b, c$. 
 In $d$ dimensions, this relation is given by \citep{Buchel:2009sk}
 \begin{eqnarray}
t_2 &=& 2\frac{(d+1)}{d}\frac{((d-2)(d+2)(d+1)\mathcal{A}+3d^2 \mathcal{B}-4d(2d+1)\mathcal{C})}{((d-1)(d+2)\mathcal{A}-2\mathcal{B}-4(d+1)\mathcal{C})},   \nonumber\\
t_4 &=& -\frac{(d+1)}{d}\frac{((d+2)(2d^2-3d-3)\mathcal{A}+2d^2(d+2) \mathcal{B}-4d(d+1)(d+2)\mathcal{C})}{((d-1)(d+2)\mathcal{A}-2\mathcal{B}-4(d+1)\mathcal{C})}                            ,
\end{eqnarray}
where ${\cal A}, {\cal B}, {\cal C}$ are linearly related to $a, b, c$ by (\ref{calabcd}) . 
Thus we have 
\begin{eqnarray}\label{t2t4abc}
 t_2 &=& \frac{2 (1+d) (-d (c-3 b d+2 c d)+a (-1+d) (4+d (8+d)))}{d \left(-2 b-c (1+d)+a \left(-6+d+d^2\right)\right)}, \nonumber\\
 t_4 &=& \frac{(1+d) (2+d) \left(d (c-2 b d+c d)+3 a \left(1+d-2 d^2\right)\right)}{d \left(-2 b-c (1+d)+a \left(-6+d+d^2\right)\right)}.
\end{eqnarray}
We can use  (\ref{t2t4abc}) and 
re-write $\delta G_R(0)$  in terms of $t_2, t_4$ which results in 
\begin{eqnarray} \label{bounds}
\delta G_R(0) 
= \left(\frac{(-1+d) d}{2 (1+d)}+\frac{(3-d) t_2}{2 (-1+d)}+\frac{\left(2+3 d-d^2\right) t_4}{(-1+d) (1+d)^2}\right)P .
\end{eqnarray} 
In this form of the sum rule and from the result
of the previous section,  it is immediately clear that   for $d>3$ 
theories which admit 
a holographic dual of 
Einstein gravity in $AdS_{d+1}$  lie at the origin in the $t_2, t_4$, i.e. at  $t_2= t_4=0$. 
However as expected for $d=3$, we have only one condition $t_4=0$.  This reflects the 
fact that there are only $2$ independent parity even constants which 
determine the three point function of the stress tensor in $d=3$ \cite{Osborn:1993cr} as 
opposed to $3$ for $d>3$. 
Therefore the fact that we obtained that the coefficient of $t_2$ vanishes
in $d=3$ is a consistency check on our derivation of the sum rule.

 We have seen that evaluation of the sum rule 
 in a $2$ derivative theory of gravity reduces to (\ref{holsum}). 
 This implies that that a necessary condition for any conformal field theory to admit 
 a Einstein gravity dual is that the shear sum rule must satisfy   (\ref{holsum}). 
It is interesting to examine if this necessary condition on the sum rule 
for Einstein gravity dual, that is 
\begin{equation}  \label{nesconst}
\delta G_R(0) =  \frac{d(d-1)}{2(d+1)}  P, 
\end{equation}
is independent of the  necessary constraint of the equality of the $\hat a = \hat c$ central charges  for  $d=4$ \footnote{We denote these central charges as $\hat a, \hat c$ to distinguish 
them from the constants $a, c$.}. 
The equality of these central charges result in \cite{Hofman:2008ar},. 
\begin{eqnarray}
\frac{\hat a}{\hat c} = \frac{9a - 2 b  - 10c}{ 3 ( 14 a - 2b - 5 c) } = 1.
\end{eqnarray}
 This implies that we have the relation 
 \begin{equation}\label{aeqc}
 33a - 4b - 5c = 0 .
 \end{equation}
 We have assumed that $ 14 a - 2b - 5 c \neq 0$. 
 Now let us examine constraint given by the condition  (\ref{nesconst}) for $d=4$. 
 We obtain  the linear relation
 \begin{equation}
 -244 a+68 b-55 c = 0.
 \end{equation}
 which is independent of (\ref{aeqc}). 
 
\begin{figure}[!htb]
  \centering
   \hspace*{-2cm}\includegraphics[scale=0.5]{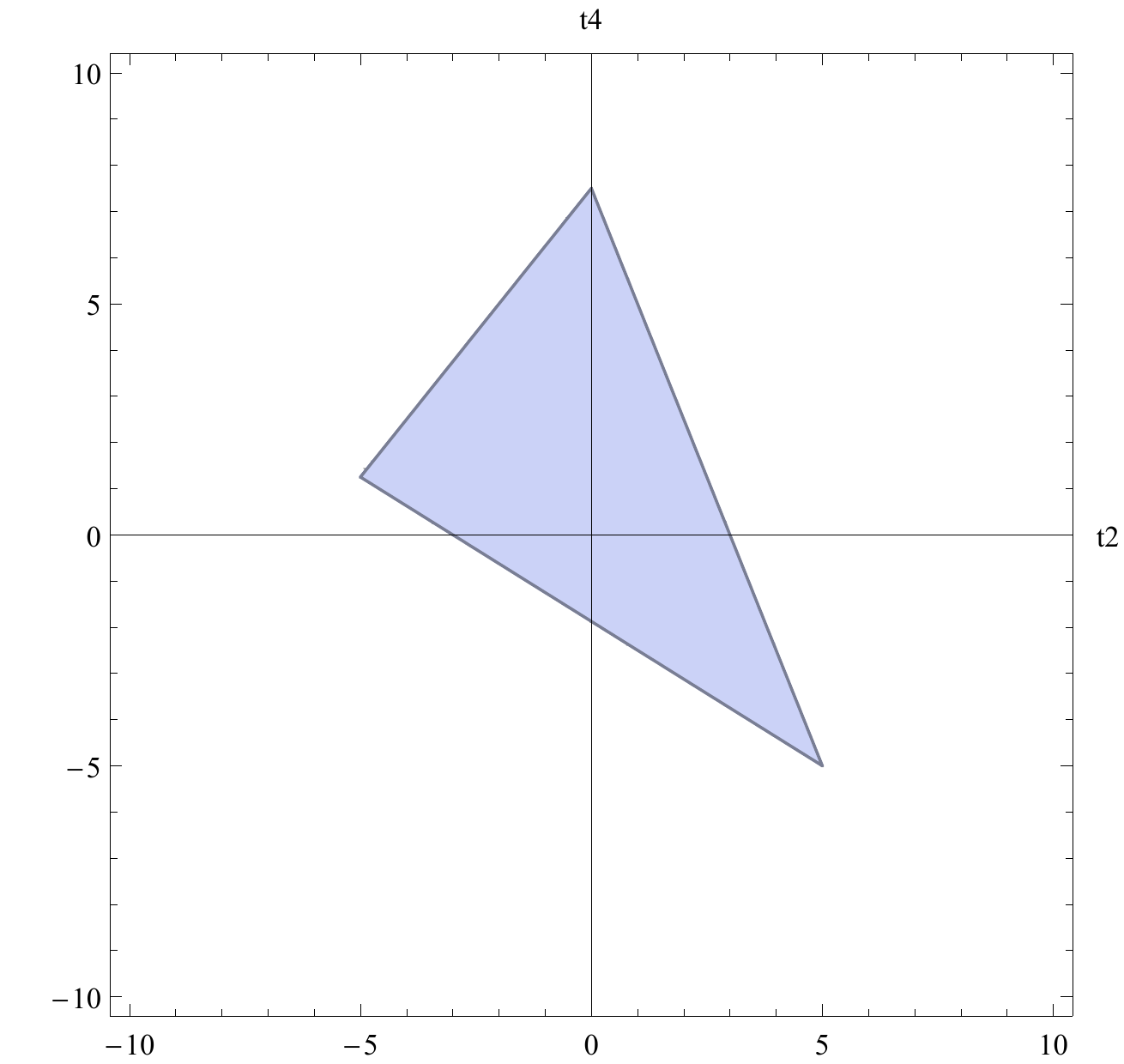} 
   \caption{The allowed domain for  $t_2$ and $t_4$ in $d=4$ 
   conformal field theories.}\label{bounddiag}
\end{figure}

 Finally let us examine the implications of the positivity of energy flux  \cite{Hofman:2008ar}constraints on the sum rule. These constraints have been recently shown to be related 
 to causality and unitarity of the conformal field theory \cite{Hartman:2015lfa,Komargodski:2016gci,Hofman:2016awc}. 
 For conformal field theories in $d>3$ dimensions, positivity of energy implies the following bounds 
 on the parameters $t_2, t_4$ \cite{Camanho:2009vw}
 \begin{eqnarray}\label{constraints}
&&1-\frac{1}{d-1}t_2-\frac{2}{(d+1)(d-1)}t_4 \geq 0, \\
&&1-\frac{1}{d-1}t_2-\frac{2}{(d+1)(d-1)}t_4+\frac{1}{2}t_2 \geq 0, \nonumber\\
&& 1-\frac{1}{d-1}t_2-\frac{2}{(d+1)(d-1)}t_4+\frac{d-2}{d-1}(t_2+t_4) \geq 0. \nonumber
\end{eqnarray}
 These bounds  imply that  conformal field theories which obey the positivity of energy/
 causality constraints lie  in the 
  region bounded by $3$ lines in the $t_2, t_4$ plane. 
  For $d=4$, this region is shown as the shaded triangle in \ref{bounddiag}. 
 Now given these inequalities in (\ref{constraints}) we can obtain bounds on the 
 function $\delta G_R(0)$ given in (\ref{bounds}). 
 The  possible extrema of  $\delta G_R(0)$ will lie on the $3$ vertices. This leads us
 to the following bounds on the sum rule 
 \begin{eqnarray}\label{causalityboundseven}
 \frac{1}{2}P \leq  \delta G_R(0) \leq \frac{d}{2} P. 
 \end{eqnarray}
 Thus the sum rule for any conformal field theory in $d>3$  obeying the 
 Causality bounds is constrained to lie between $\frac{P}{2} $ and  and $\frac{d}{2}P$. 
 
 For $d=3$, ignoring the parity odd term 
 in the three point function of the stress tensor  found by  \cite{Giombi:2011rz}, 
 it can be seen that the first inequality 
holds reduces to an equality \cite{Komargodski:2016gci}.   Therefore we obtain
\begin{equation}
t_2  = 2 - \frac{t_4}{2}.
\end{equation}
The remaining two inequalities reduce to 
\begin{equation}
-4 \leq t_4 \leq 4.
\end{equation}
This implies that the sum rule for parity even conformal field theories in $d=3$ is 
constrained to lie between
\begin{equation}\label{3dcaus}
\frac{P}{2}  \leq \delta G_R(0) \leq  P.
\end{equation}

\section{Applications}\label{applications}

The parameters $a,b$ and $c$ completely specifies the three 
point function of the stress tensor  in a  conformal field theory. 
In this section we will evaluate $\delta G_R(0)$ using the expression 
in (\ref{sumabc}) for well studied examples of conformal field theories
in $d=3, 4, 6$ dimensions. 
All the theories that we consider here except that of Chern-Simons theory coupled to fundamental 
fermions have  gravitational duals and are supersymmetric. 
 We will show that for the supersymmetric theories 
considered here which admit a gravity dual, the coefficient in $\delta G_R(0)$
evaluated at 
 at weak coupling of these 
theories agrees 
precisely with that in gravity.  This strongly suggests that the sum rule is not 
renormalized for these theories. 

\subsection{$d=3$}
Substituting for $d=3$ in the sum rule (\ref{sumabc}) we find that it takes the 
form 
\begin{eqnarray} \label{3dsum}
\delta G_R(0) &=&\frac{P (-13 a-9 b+2 c)}{6 a-2 (b+2 c)}.
\end{eqnarray}
Free conformal field theories in $d=3$ consist of $n_s$ real bosons and 
$n_f$ Dirac fermions.  The contribution of these fields to the constants
$a, b, c$ are given by \cite{Osborn:1993cr}
\begin{eqnarray}\label{3dabc}
a = \frac{27 n_s}{4096 \pi ^3} , \qquad
b =-\frac{9 (8 n_f+9 n_s)}{4096 \pi ^3}, \qquad
c = -\frac{9 (16 n_f+ n_s)}{4096 \pi ^3}.
\end{eqnarray}
A well studied example of a theory in $d=3$  which admits $AdS_4$ as its 
gravity dual 
is that of the M2-brane. 
While the theory of interacting multiple M2-branes is not known, the 
field content of a single M2-brane is known to consist of 
$8$ real scalars, and 8 Majorana fermions which are  equivalent to 
$4$ Dirac fermions. 
The contribution of this  field content to $a, b, c$ is given by  
\begin{eqnarray}\label{m2abc}
a = \frac{27}{8(4\pi)^3}, \qquad
b  =  \frac{-117}{8(4\pi)^3}, \qquad 
c = \frac{-81}{8(4\pi)^3}. 
\end{eqnarray}
Evaluating the sum rule we obtain 
\begin{equation}
\delta G_R(0)|_{\rm M2} = \frac{3P}{4}.
\end{equation}
This value for the sum rule 
precisely agrees with that obtained in gravity for $d=3$  given in (\ref{holsum}). 

Another theory which admits $AdS_4$ as its gravity dual is the ABJM theory
\cite{Aharony:2008ug}. 
The values of $a, b, c$ for the interacting theory is not known. 
However at weak coupling, the theory consists of 
$3$ sets of $8$ real scalars  with $N^2$ internal components and 
$3$ sets of $4$ Dirac fermions with $N^2$ internal components
\footnote{The $U(N) \times U(N)$ Chern-Simons field is not dynamical.}.
Thus  the values of $a, b, c$ for the ABJM theory is $3 N^2 $ times that 
of the M2-brane given in (\ref{m2abc}). 
The sum rule (\ref{3dsum}) is given by the ratio  of these constants 
and   remains the same as that of the M2-brane. 
Thus for the ABJM theory,  the sum rule at weak coupling agrees precisely 
with the result at strong coupling.  
\begin{equation}
\delta G_R(0)|_{\rm ABJM} = \frac{3P}{4}.
\end{equation}

\subsection*{Large $N$ Chern-Simons vector theories }

Recent work of \cite{Giombi:2011kc,Aharony:2011jz,Aharony:2012nh,GurAri:2012is} have shown that 
the planar limit of 
$U(N)$ Chern-Simons theories  at level $k$ 
coupled to fermions or bosons in the fundamental representation
are solvable in the large $N$ limit.  Let us first restrict to the case of 
the fermionic theory. 
Using the analysis of \cite{MZ}, it can be seen that the
the three point function of the stress tensor of the interacting theory  can be written as
\begin{equation}
\langle TTT\rangle_{\rm int\, fermion} 
= n_s(f) \langle TTT\rangle_{\rm free\, boson}
+ n_f(f) \langle TTT\rangle_{\rm free\, fermion}
+ \gamma (f)\langle TTT \rangle_{\rm parity\, odd},
\end{equation}
where
\begin{eqnarray}
& & n_s(f) = 2N\frac{\sin \theta}{ \theta} \sin^2 \frac{\theta}{2}, \qquad 
 n_f(f) = 2N\frac{\sin \theta}{ \theta} \cos^2 \frac{\theta}{2}, \\ \nonumber
& & c(f) = N \frac{\sin^2 \theta}{\theta} , \qquad 
\theta = \frac{N_f}{k}.
\end{eqnarray}
The $f$ in the brackets refers to the fact that the theory consists of fundamental 
fermions and a summary of the derivation of this result is given in 
appendix \ref{sec:intro}. The parity odd term does not play any role in the shear channel. 
Therefore we can treat the theory as a theory of $n_s(f)$ free real scalars and 
$n_f(f)$ free real fermions  and therfore $\frac{n_f(f)}{2}$ complex fermions. Evaluating the parameters $a, b, c$ using (\ref{3dabc}) 
we obtain 
\begin{eqnarray}\label{largenabc}
a &=& \frac{27 N \sin ^2\left(\frac{\theta }{2}\right) \sin (\theta )}{2048 \pi ^3 \theta },  \\
b &=& \frac{9 N \sin (\theta ) (5 \cos (\theta )-13)}{4096 \pi ^3 \theta }, \nonumber\\
c &=&-\frac{9 N \sin (\theta ) (7 \cos (\theta )+9)}{4096 \pi ^3 \theta }.\nonumber\\
\end{eqnarray}
Using these values in the sum rule (\ref{3dsum}) we obtain 
\begin{equation}
\delta G_R(0) = -\frac{1}{4} P (\cos \theta -3).
\end{equation}
First note that the the causality bounds for the sum rule  given in 
(\ref{3dcaus}) is satisfied as $\theta$ is dialled from $\theta =0, \cos \theta = 1 $ the theory of 
free fermions  to $\theta = \pi, \cos \theta = -1$. 
At $\theta = \pi$, the theory of free bosons. 
Another  interesting observation from this result for the sum rule is that
at $\cos \theta = 0$, the result for the sum rule 
agrees with  that obtained from Einstein's theory in $AdS_4$. 
The is because  the theory at this point ,  
effectively consists of $n_f$ free complex fermions and $n_s$ real scalars with 
$\frac{n_s}{n_f} = 2$. In fact it can be seen that for any theory which 
satisfies $\frac{n_s}{n_f} = 2$ the sum rule agrees with that 
in gravity. 
The M2-brane theory as well as the ABJM theory and other related theory which 
admit a gravity dual satisfies the condition $n_s/n_f = 2$.
It will be interesting to study if there is any simplification for the dual higher 
spin Vasiliev theory at $\theta =  \frac{\pi}{2}$.

\subsection{$d=4$}

In $d=4$ The sum rule in (\ref{sumabc}) takes the following form for $d=4$.
\begin{eqnarray}\label{4dsum}
\delta G_R(0) &=&\frac{P (5 c-16 (2 a+b))}{14 a-2 b-5 c}.
\end{eqnarray}
The values of $a, b, c$ for a theory consisting of free  $n_s$ real scalars, $n_f$ Dirac fermions and 
$n_v$ vectors  is given by 
\begin{eqnarray} \label{4dabc}
&& a = \frac{1}{27\pi^6} ( n_s - 54 n_v) , \qquad b = - \frac{1}{54\pi^6} ( 8 n_s + 27 n_f) , 
\\ \nonumber
& & c = -\frac{1}{27\pi^6} ( n_s +  27( n_f + 8 n_v) ).
\end{eqnarray}
Consider the  case of ${\cal N}=4$ super Yang-Mills which consists of 
6 real scalars, 4 Majorana fermions which is equivalent to 2 Dirac fermions
and 1 vector each in the adjoint representation of the gauge group $SU(N_c)$.
Substituting these values for $n_s, n_f$ and $n_v$ in to (\ref{4dabc}) we obtain
\begin{eqnarray}
a=-\frac{16}{9\pi^6}(N_c^2-1),\qquad
b=-\frac{17}{9\pi^6}(N_c^2-1),\qquad
c=-\frac{92}{9\pi^6}(N_c^2-1).
\end{eqnarray}
With these values the sum rule (\ref{4dsum}) reduces to 
\begin{equation}
\delta G_R(0) = \frac{6P}{5}.
\end{equation}
which precisely agrees with the gravity result (\ref{holsum}) for $d=4$.
This result  that the sum rule at weak coupling 
for ${\cal N}=4$ Yang-Mills agrees with that in gravity 
was also observed by \cite{Romatschke:2009ng}.

As another consistency check for the bounds on the sum rule  we have obtained in 
(\ref{causalityboundseven}) let us examine the case of 
free $SU(N)$ Yang-Mills  in $d=4$. The RHS of the shear sum rule 
was evaluated in \cite{Romatschke:2009ng} and was shown to be 
$2P$ which saturates the upper bound bound  in $d=4$.

\subsection{$d=6$}

In $d=6$, the free conformal field theories consist of 
$n_s$ real scalars,  $n_f$ Dirac fermions and $n_t$,  rank 2-forms. 
In  such a theory, the values of $a, b, c$ are given by 
\begin{eqnarray}
a &=&\frac{27 (-250 n_t+n_s)}{125 \pi ^9}, \\
b &=&-\frac{18 (125 (6 n_t+n_f)+9 n_s)}{125 \pi ^9}, \nonumber\\
c &=& -\frac{9 \left( 125( 96 n_t+8 n_f) +16 n_s \right)}{250 \pi ^9}.\nonumber
\end{eqnarray}
The contribution of real scalars and Dirac fermions to the constants
$a, b, c$ can be obtained from  \cite{Osborn:1993cr}, while the contribution of
self dual tensors has been evaluated in  \cite{Buchel:2009sk}. 
The sum rule in $d=6$ takes the   form,
 \begin{eqnarray}\label{6dsum}
\delta G_R(0)  &=& -\frac{2 P (56 a+18 b-7 c)}{36 a-2 b-7 c}.
 \end{eqnarray}
 The most studied  theory in $d=6$ is that of the $M5$ brane which admits a holographic
 dual. While the theory of multiple  M5-branes is not known, we can consider the
 theory of a single M5-brane whose field content consists of the $(2, 0)$ tensor multiplet
 which is made up of a single self dual tensor $n_t = 1/2$,  2 Weyl fermions which is equivalent
 to a single Dirac fermion and 5 real scalars. 
 Using this field content   we obtain 
 \begin{eqnarray}
a = -\frac{2592}{100 \pi ^9}, \qquad 
b = -\frac{7848}{100 \pi ^9}, \qquad 
c = -\frac{25488}{100 \pi ^9}. 
\end{eqnarray}
Substituting these values into the sum rule (\ref{6dsum}) we obtain 
\begin{equation}
\delta G_R(0) = \frac{15 P}{7}.
\end{equation}
Again, the agrees with the result from gravity for $d=6$ given in 
(\ref{holsum}). 
\subsection{Gauss Bonnet Gravity}

As a simple consistency check, we  will now verify that the 
general bounds derived for the sum rule in (\ref{causalityboundseven}) 
is consistent with the existing bound on the coefficient of the coupling of the 
Gauss-Bonnet term in $AdS_{d+1}$ with $d\geq 4$. 
Bounds for the Gauss-Bonnet coupling $\lambda_{GB}$ were obtained 
in \citep{Buchel:2009sk} using the causality bounds of Maldacena and Hofman given in 
in (\ref{constraints}). 
 From the analysis of \citep{Buchel:2009sk} we obtain 
\begin{eqnarray}
t_4 &=& 0 \nonumber\\
t_2 &=& \frac{4 f_\infty \lambda_{GB}}{1-2f_\infty \lambda_{GB}} \frac{(d)(d-1)}{(d-2)(d-3)}
\end{eqnarray}
where,
\begin{eqnarray}
f_\infty &=& \frac{1-\sqrt{1-4\lambda_{GB}}}{2\lambda_{GB}}
\end{eqnarray}
Now causality bounds  in (\ref{constraints}) restrict the Gauss-Bonnet  term to lie 
within
\begin{eqnarray}\label{bgb}
-\frac{(3d+2)(d-2)}{4(d+2)^2} \leq \lambda_{GB} \leq \frac{(d-2)(d-3)(d^2-d+6)}{4(d^2-3d+6)^2}
\end{eqnarray}
Let us  verify that within this window of the Gauss-Bonnet coupling, the 
bound for the sum rule  in \ref{causalityboundseven} is satisfied. 
Substituting the values of $t_2, t_4$ for Gauss-Bonnet gravity in 
(\ref{bounds}) we obtain 
\begin{eqnarray}
\delta G(0)&=& 
\left( \frac{d(d-1)}{2(d+1)} + d\frac{\left( 1- \frac{1}{\sqrt{1-4\lambda_{GB}}} \right)}{(d-2)} 
\right) P
\end{eqnarray}
The bounds on the Gauss-Bonnet coupling in (\ref{bgb}) 
imply that 
\begin{eqnarray}
\left( \frac{1}{2} + \frac{1}{d+1} \right ) P\leq \delta G_{GB}(0) \leq 
\left( \frac{d}{2} - \frac{d-1}{2(d+1)} \right) P
\end{eqnarray}
It is easy to see that this is within the general bound derived for the sum rule 
in (\ref{causalityboundseven}). 
This result  therefore  serves as a minor consistency check on the coefficient of 
$t_2$ in (\ref{bounds}) 
since  $t_4$ is vanishing in Gauss-Bonnet gravity.

\section{Retarded Greens function in other channels}\label{other}

In this section we study the high frequency behavior of the Greens function in the 
vector and the sound  channels. 
We will  Fourier transform the OPE in these channels and obtain the finite 
term.  After factorizing   the appropriate tensor structure  in these channels we 
show that the finite term  in these channels contains 
the Hofman-Maldacena coefficients 
\begin{eqnarray}  \label{hofmalc12}
a_{T, 1} &=& \frac{1}{8} \frac{d (b (2-3 d)+2 c d-a (-8+d (6+d))) P }{\left(- (2 b+c+c d)+ a \left(-6+d+d^2\right)\right) }, 
\\ \nonumber
a_{T, 2} &=& -\frac{1}{32}
\frac{ (4 a+2 b-c) (-2+d) d   }{\left(-2 b-c (1+d)+a \left(-6+d+d^2\right)\right)}.
\end{eqnarray}

\subsection{The vector channel}

The Greens function and its Fourier in this  channel is  defined by the 
correlator 
\begin{eqnarray}
G_{R; V} (t,x) &=& i \theta(t)\langle[T_{xt},T_{xz}]\rangle ,\\
G_{R; V} (\omega, p_z) &=& \int d^dx e^{i\omega t - i p_z. z} i \theta(t)\langle[T_{xt},T_{xz}]\rangle. \nonumber
\end{eqnarray}        
As argued in the case of  the shear channel, the high frequencybehavior
of this Greens can be obtained by studying the OPE of the stress tensors
in these channels. 
In the appendix \ref{fouriertransformvector} we have evaluated the Fourier transform of the 
OPE coefficient $\hat A_{xtxz\alpha\beta} ( \omega, p_z) \langle T_{\alpha\beta} \rangle$.
The result is given by 
\begin{eqnarray}
\hat A_{x\tau xz} (\omega, p_z) 
&=& -\frac{ p_z \omega}{(p_z^2+\omega^2)} G_2(\omega, p_z), 
\end{eqnarray}
where 
\begin{eqnarray}\label{defg2}
G_2(\omega, p_z) = - \delta G_R(0)  +    a_{T, 1}\frac{ 8 d p_z^2}{\omega^2 + p_z^2} P.
\end{eqnarray}
$\delta G_R(0)$ is the RHS of the shear sum rule defined in (\ref{sumabc}) 
and $a_{T, 1}$ is the Hofman-Maldacena coefficient in the vector channel defined  as
\begin{equation}
a_{T, 1} = \frac{1}{8} \frac{ b (2-3 d)+2 c d-a (-8+d (6+d))  }{- (2 b+c+c d)+ a \left(-6+d+d^2\right) }.
\end{equation} 
It is indeed interesting that both the RHS of the shear sum rule as well 
as the 2nd Hofman-Maldacena coefficient appears in the high frequencybehavior 
in this channel. 
Further more  note from (\ref{defg2}), that starting from the term $(\frac{p_z}{\omega})^2$,
the expansion of $G_2(\omega, p_z)$ is entirely determined by the 
Hofman-Maldacena coefficient $a_{T,1}$. 
Expressed in terms of $t_2$ and $t_4$, this takes the form:
\begin{equation}
    G_2(\omega, p_z) =- \delta G_R(0)   +\left(-\frac{4 (d-3) 
   t_2}{\left(d^2-1\right)}+\frac{16 
   t_4}{(d-1) (d+1)^2}-\frac{8  }{(d+1)}\right)\frac{d p_z^2}{p_z^2+\omega^2} P
\end{equation}
For $d=3$, we see that the expression is independent of $t_2$.

For theories with a gravity dual, this becomes
\begin{equation}
    G_2(\omega, p_z) = -d \left( \frac{d-1}{2(d+1)} + \frac{8}{d+1}\frac{p_z^2}{p_z^2+\omega^2}\right) P,
\end{equation}
which agrees with the supersymmetric examples we considered in section 5. 
For parity odd Chern-Simons theory coupled to fundamental matter, we expect an 
additional contribution from the parity-odd term in the three-point function of the stress tensor.

\subsection{The sound channel}

We now examine the sound channel  in which the retarded Greens function 
and its Fourier transform is defined by 
\begin{eqnarray}
G_{R;S} (t,x) &=& i \theta(t)\langle[T_{tt},T_{tt}]\rangle \\ \nonumber
G_{R;S} (\omega, p_z) &=& \int d^dx e^{i\omega t - i p_z   z} i \theta(t)\langle[T_{tt},T_{tt}]\rangle
\end{eqnarray} 
The momentum is along any spatial direction. 
Again we can resort to the OPE to study the high frequency behavior. 
In the appendix (\ref{fouriertransformscalar}) we Fourier transform the OPE coefficient
corresponding to this channel and obtain 
\begin{eqnarray}
 \hat A_{tttt\alpha\beta} (\omega, p_z)  \langle T_{\alpha\beta} \rangle 
 &=& \frac{p_z^4}{(p_z^2+\omega^2)^2}G_3( \omega, p_z).
\end{eqnarray}
The function $G_3(\omega, p_z) $ admits an expansion 
which is given by 
\begin{eqnarray}
G_3( \omega, p_z) &=& \left(  F_1 ( \frac{\omega}{p_z}  )^4  
+ F_2 ( \frac{\omega}{p_z}  )^2 + F_3 +   a_{T, 2}  \frac{  64(\frac{ p_z}{\omega})^2  }
{ 1 + (\frac{p_z}{\omega})^2} \right) P.
\end{eqnarray}
Here $F_1, F_2, F_3$ are ratios of linear functions of the constants $a, b, c$. 
They can be written in the variables $t_2, t_4$. 
However what is interesting is that again
starting from the term $ (\frac{ p_z}{\omega})^2 $  the entire expansion 
is determined by 
$a_{T, 2}$, the Hofman-Maldacena coefficient in the tensor channel
which is given by 
\begin{equation}
a_{T, 2} = -\frac{1}{32}
\frac{ (4 a+2 b-c) (-2+d) d   }{\left(-2 b-c (1+d)+a \left(-6+d+d^2\right)\right)}.
\end{equation}

While the expressions  for $F_1, F_2, F_3$  in terms of $a, b, c$ or $t_2, t_3$ in arbitrary 
dimensions are fairly lengthy, we can present them 
for theories in $d=3$, $d=4$ and $d=6$ with a gravitational dual. 
We note that, while $a_{T,2}$ is independent of $t_2$ for $d=3$, some of the 
remaining terms in the expression are not independent of $t_2$ for $d=3$. 

For $d=3$, we set $t_2=2$ and $t_4=0$, and obtain:
\begin{eqnarray}
G_3 \Big|_{d\rightarrow 3} &=& \left(\frac{32}{3} \pi   -\frac{151}{18}\right)\frac{\omega^4}{p_z^4} P +\left({\frac{64}{3} \pi 
  -\frac{115 }{12}}\right)\frac{\omega^2}{p_z^2}P+\left(\frac{32 \pi 
   P}{3}-\frac{P}{36}\right) \nonumber \\
   & & \qquad +\frac{3 p_z^2 P}{8
   (\omega ^2+p_z^2)}
\end{eqnarray}
For $d=4$ and $d=6$, we set $t_2=t_4=0$, and obtain:
\begin{eqnarray} \nonumber
G_3 \Big|_{d\rightarrow 4} & = & \left({6 \pi ^2 -\frac{527}{10}}\right)\frac{ \omega ^4}{p_z^4}P 
+
\left({12 \pi ^2-\frac{3401 }{60}}\right)\frac{\omega^2}{p_z^2}P+\left(6
   \pi ^2-\frac{131}{60}\right)P
   +\frac{16
   p_z^2 P}{45 (\omega ^2+p_z^2)} 
   \\ \nonumber
   G_3 \Big|_{d\rightarrow 6} & = & \left({\frac{10}{3} \pi ^3 -\frac{8552 }{63}}\right)\frac{\omega
   ^4}{p_z^4}P+\left({\frac{20}{3} \pi ^3-\frac{15046 }{105}}\right)\frac{\omega^2}{p_z^2}P+
   \left(\frac{10 \pi ^3}{3}-\frac{1598 }{315}\right)P \\ 
   && \qquad +\frac{48 p_z^2 P}{175 (\omega ^2+p_z^2)}
\end{eqnarray}
These results agree with the supersymmetric examples of section \ref{applications}.

\section{Conclusions}\label{conclusion}

We have derived the shear sum rule obeyed by conformal field theories
in dimensions $d>0$. 
Assuming analyticity in the frequency plane, 
 the  sum rule holds when there are  no operators of dimension 
$\Delta \leq d$ gain expectation value in the thermal vacuum. 
The RHS of the sum rule is determined by constants $a, b, c$ of the
three point function of the stress tensor, these can be written in terms
of the Hofman-Maldacena variables $t_2, t_4$. 
We showed that for theories that admit a dual description in terms of  Einstein gravity 
in $AdS_d$, the  RHS of the sum rule reduces to 
$\frac{d}{2(d+1)} \epsilon$. We have also determined bounds 
on the sum rule using the causality constraints on $t_2, t_4$. 
One interesting observation in the derivation of the sum rule is that the 
high frequency expansion of the Greens function in the shear channel 
is determined by the Hofman-Maldacena coefficient $a_{T, 0}$ given in (\ref{malcoef0}).  

The shear sum rule given in (\ref{bounds}) was cross checked 
by evaluating the retarded Greens function  by considering 
 a minimally coupled scalar in $AdS_{d+1}$. As one can see, this analysis 
 only tests the coefficient independent of $t_2, t_4$ in the sum rule. 
 Note that  the vanishing of the coefficient of $t_2$  for $d=3$ in (\ref{bounds}) 
 is consistent with the fact that that there is only $2$ independent 
 parameters determining the three point function of stress tensor in 
 $d=3$.  
 Finally the fact that for Gauss-Bonnet gravity we have shown that the 
 shear sum rule lies  within  the general bounds predicted in (\ref{causalityboundseven})
 is also a minor consistency check on the coefficient of $t_2$ since 
 $t_4$ is vanishing in Gauss-Bonnet gravity. 
 However it will be useful to directly check both the coefficient of $t_2$ and $t_4$
 by evaluating the retarded Greens function in  higher derivative theories of gravity. 
 We hope to report on this in the near future.

We have also studied the high frequency expansion of the retarded Greens function 
in the vector and sound channels.  We observed that again in these channels, 
the high frequency expansion is determined by Hofman-Maldacena coefficients  $a_{T, 1}, 
a_{T, 2}$  given in (\ref{hofmalc12}) 
for the vector and sound channels respectively. 
It will be interesting to cast this observation as sum rules in these channels 
and perform similar consistency checks using holography as done 
in this paper for the shear channel. 
The observation of  the appearance of 
Hofman-Maldacena coefficients in the OPE of stress tensors
have also been made in \cite{Komargodski:2016gci} in the kinematic regime tuned to study 
deep inelastic scattering. Here the OPE is taken on an one particle state of the theory. 
It will be interesting to relate this to 
the study done in this paper where the OPE is taken in the  thermal vacuum.

Finally this work points out that interesting and useful constraints on 
the spectral density  can be obtained using conformal invariance and causality. 
It will be rewarding  to explore this direction further. 

\acknowledgments

We wish to thank R. Loganayagam for  discussions and questions which led us to explore the 
constraints of conformal invariance on spectral densities. 
We thank Zohar Komargodski for  very useful correspondence, questions 
and for a very careful  reading of the manuscript. 
We  thank William Witczak-Krempa for correspondence which highlighted 
the role of marginal and relevant operators in the CFT. 
We also thank   Gautam Mandal, 
 Shiraz Minwalla, Suvrat Raju, Ashoke Sen, Aninda Sinha and Sandip
Trivedi for useful comments at presentations of this work  in  seminars and discussions
during  the course   of this project.

\appendix 

\section{Fourier transform of the OPE }\label{append}

This appendix consists of $3$ sub-sections each of which provides the details 
of the Fourier transform of the tensor structure  of the coefficient 
$\hat A_{\mu\nu\rho\sigma\alpha\beta}$ of the OPE of the stress tensor in 
$d>2$ dimensions given in (\ref{tenstu}). 
Section \ref{fouriertransformtensor} deals with the Fourier transform in the 
shear channel. Here the indices involved in the OPE of the two stress tensors
are spatial and orthogonal to each other, the OPE considered 
is $\langle T_{xy} T_{xy}\rangle $. 
Section \ref{fouriertransformvector} evaluates the Fourier transform in the vector channel where 
there is one spatial index common in the stress tensor and one 
time direction in of the stress tensor. This OPE channel is given by 
$\langle T_{xt} T_{xz}\rangle $. 
Finally in section \ref{fouriertransformscalar} 
we evaluate the Fourier transform in the sound channel given by 
$\langle T_{tt} T_{tt} \rangle$. In all these channels, the momentum will be taken in the 
$z$ direction which is orthogonal to the $x, y$ direction. 
However our results for the Fourier transform can be easily  extended to 
momenta in all directions, since we have performed the basic integrals 
required in section \ref{integrals} with momenta turned on in all directions.

\subsection{The shear channel} \label{fouriertransformtensor}

Let us consider the OPE coefficient $\hat A_{xyxy\alpha\beta}$. 
We will finally require only the coefficients $\alpha = \beta$, since we take
expectation values in the thermal vacuum. 
From \cite{Osborn:1993cr}, this tensor structure is given by 
\begin{eqnarray}\label{expansion1}
&&\hat{A}_{\mu\nu\rho\sigma\alpha\beta} C_T = \frac{(d-2)}{d+2}(4a+2b-c)H^1_{\alpha \beta \mu\nu\rho\sigma}(s) +  \frac{1}{d}(da+b-c) H^2_{\alpha\beta \mu\nu\rho\sigma}(s) \nonumber\\
&&-\frac{d(d-2)a-(d-2)b-2c}{d(d+2)} (H^2_{\mu\nu\rho\sigma\alpha\beta}(s)+H^2_{\rho\sigma\mu\nu\alpha\beta}(s))+\frac{2da+2b-c}{d(d-2)} H^3_{\alpha\beta \mu\nu\rho\sigma}(s)\nonumber\\
&&-\frac{2(d-2)a-b-c}{d(d-2)}H^4_{\alpha\beta \mu\nu\rho\sigma}(s)-\frac{2((d-2)a-c)}{d(d-2)}(H^3_{\mu\nu\rho\sigma\alpha\beta}(s)+H^3_{\rho\sigma\mu\nu\alpha\beta}(s))\nonumber\\
&&+\frac{((d-2)(2a+b)-dc)}{d(d^2-4)}(H^4_{\mu\nu\rho\sigma\alpha\beta}+H^4_{\rho\sigma\mu\nu\alpha\beta})(s)\nonumber\\
&&+(C h^5_{\mu\nu\rho\sigma\alpha\beta}+D(\delta_{\mu\nu} h^3_{\rho\sigma \alpha \beta}+ \delta_{\rho \sigma} h^3_{\mu \nu \alpha \beta}))S_d \delta^d(s),\nonumber\\
&=& I_1+I_2+I_3+I_4+I_5+I_6+I_7+I_8,
\end{eqnarray}
where,
\begin{eqnarray} \label{expansion2}
H^1_{\alpha \beta xyxy}(s)&=&(\partial_\alpha \partial_\beta - \frac{1}{d} \delta_{\alpha \beta} \partial^2)\left(\frac{x^2y^2}{(t^2+x^2+y^2+z^2+x_1^2+\cdots + x_{d-4}^2)^\frac{d+2}{2}}\right),\nonumber\\
H^2_{\alpha\beta xyxy}(s)&=&(\partial_x^2+\partial_y^2)\frac{1}{(t^2+r^2)^\frac{d-2}{2}}(\frac{s_\alpha s_\beta}{t^2+r^2}-\frac{1}{d} \delta_{\alpha \beta}),\nonumber\\
H^3_{\alpha\beta xyxy}(s)&=&(\partial_\alpha \partial_\beta-\frac{1}{d}\delta_{\alpha\beta} \partial^2) \frac{1}{(t^2+r^2)^\frac{d-2}{2}},\nonumber\\
H^4_{\alpha\beta xyxy}(s)&=&((2\delta_{\alpha x}\delta_{\beta y}-\frac{2}{d}\delta_{\alpha \beta})\partial_y^2 +(2\delta_{\alpha y}\delta_{\beta y}-\frac{2}{d}\delta_{\alpha \beta})\partial_x^2) \frac{1}{(t^2+r^2)^\frac{d-2}{2}},\nonumber\\
H^3_{xyxy\alpha\beta}(s)&=&(\delta_{x\alpha}\delta_{y\beta}+\delta_{y\alpha}\delta_{x\beta})\partial_x \partial_y \frac{1}{(t^2+r^2)^\frac{d-2}{2}},\nonumber\\
H^4_{xyxy\alpha\beta}(s)&=&(2\delta_{y\alpha}\partial_y\partial_\beta +2\delta_{x\alpha}\partial_x\partial_\beta-\frac{2}{d}\delta_{\alpha\beta}(\partial_y^2+\partial_x^2)) \frac{1}{(t^2+r^2)^\frac{d-2}{2}},\nonumber\\
h^5_{xyxy\alpha\beta}&=&2\delta_{y\alpha}\delta_{y\beta}+2\delta_{x\alpha}\delta_{x\beta}-\frac{4}{d}\delta_{\alpha\beta},\nonumber\\
C_T &=& \frac{8\pi^\frac{d}{2}}{\Gamma(\frac{d}{2})} \frac{(d-2)(d+3)a-2b-(d+1)c}{d(d+2)},\nonumber\\
C &=&\frac{(d-2)(2a+b)-dc}{d(d+2)}, \qquad S_d = \frac{2 \pi ^{\frac{d}{2}}}
{\Gamma(\frac{d}{2})}. \nonumber
\end{eqnarray} 
To Fourier transform these tensor structures, we will first 
Fourier transform the tensor structures on which the derivatives act. 
For example  for  $H^1(s)$ defined in \ref{expansion2} we will Fourier transform 
the expression in the curved brackets and then the action of the derivatives is
obtained by inserting the appropriate momenta. 
We consider each of the terms $I_i$ with $i= 1\cdots 8$ individually. 
For the moment, let us restrict our case for $d>3$.  This is because we can find $2$ directions $x, y$ perpendicular
to the momentum direction $z$.  
However the as we will discuss towards the end of this section  we have verified that the final expression 
for the sum rule for $d=3$ is a natural extrapolation of the result for $d>3$ to $d=3$.

\subsubsection*{Fourier transform: $\textbf{ I}_1$}

\begin{eqnarray}
I_1(s)&=&\frac{d-2}{d+2}(4a+2b-c)H^1_{\alpha \beta xyxy}(s) \langle T_{\alpha\beta} (0)  \rangle, \nonumber\\
\hat I_1(\omega, p)&=&\frac{d-2}{d+2}(4a+2b-c)H^1_{\alpha \beta xyxy}(\omega,  p)
\langle T_{\alpha\beta} (0), \rangle 
\nonumber\\
\sum_{i = 1}^{d-1} H^1_{iixyxy}(\omega,  p = p_z )&=&
\left[ \frac{(d-1) \left(  p^2 +\omega ^2\right)}{d}-  p^2 \right] \times \nonumber\\
&&\int d^dxe^{-i p z -i\omega t}
\left(\frac{x^2y^2}{(t^2+x^2+y^2+z^2+x_1^2+\cdots + x_{d-4}^2)^\frac{d+2}{2}}\right),\nonumber\\
&=&\frac{\pi ^{d/2} \left(d \omega ^2- p^2-\omega ^2\right)}
{d \Gamma \left(\frac{d}{2}+1\right) \left( p^2+\omega ^2\right)}. 
\end{eqnarray}
Here to perform the Fourier transform we have used  the integral of type 3 derived in section  \ref{inttype3}. 
Similarly the Fourier transform for the time component is given by 
\begin{equation}
H^1_{ttxyxy}(\omega, p)=-\frac{\pi ^{d/2} \left(d \omega ^2- p^2-\omega ^2\right)}
{d \Gamma \left(\frac{d}{2}+1\right) \left(p^2+\omega ^2\right)}.
\end{equation}
Thus combining these expressions along with the expectation value of 
the stress tensor in the thermal vacuum we obtain 
\begin{eqnarray}
I_1(\omega, p)&=&\frac{(d-2) (4 a+2 b-c)}{d+2} \frac{ \pi ^{d/2}
\left(d \omega ^2- p^2-\omega ^2\right) }{d \Gamma \left(\frac{d}{2}+1\right)
\left( p^2+\omega ^2\right)} ( P - \epsilon_E). 
\end{eqnarray}
\subsubsection*{Fourier transform: $\textbf{I}_2$}
\begin{eqnarray}
&& I_2(s)= \frac{1}{d}(da+b-c) H^2_{\alpha\beta xyxy}(s)\langle T_{\alpha \beta} (0) \rangle \\,
&& \hat I_2(\omega, \vec p) =  \frac{1}{d}(da+b-c) H^2_{\alpha\beta xyxy}(\omega, \vec  p)
\langle  T_{\alpha \beta}(0) \rangle,  \nonumber\\
&&H^2_{\alpha\beta xyxy}(\omega, \vec p)=(-p_x^2-p_y^2)\int d^dx e^{-i \vec p\cdot \vec r-
i\omega t}\frac{1}{(t^2+r^2)^\frac{d-2}{2}}
(\frac{s_\alpha s_\beta}{t^2+\vec r^{\, 2} }-\frac{1}{d} \delta_{\alpha \beta}),\nonumber\\
&&\sum_{i = 1}^{d-1} H^2_{ii xyxy}(\omega, \vec  p)=\left(-p_x^2-p_y^2\right) \left(\frac{2 \pi ^{d/2} 
\left((d-3) \left( \vec p^{\, 2} \right)+
(d-1) \omega ^2\right)}{\Gamma \left(\frac{d}{2}\right) 
\left( \vec p^{\, 2} +\omega ^2\right)^2}
 -\frac{(d-1) \left(4 \pi ^{d/2}\right)}
{d \left(\Gamma \left(\frac{d}{2}-1\right) \left(\vec p^{\, 2} +\omega ^2\right)\right)}\right).\nonumber
\end{eqnarray}
Here $\vec p$ refers to all the spatial directions of the momentum, similarly  $\vec r$ refers to
the spatial co-ordinate.  The Fourier transform has been performed using 
(\ref{inttype1}) and (\ref{inttype2})
Now it is clear that on taking the limit $p_x, p_y \rightarrow 0$, the above expression vanishes. 
\begin{equation}
 \lim_{p_x, p_y \rightarrow 0} \sum_{i = 1}^{d-1} H^2_{ii xyxy}(\omega, \vec  p) =0.
\end{equation}
Similarly it can be shown  that
\begin{equation}
  \lim_{p_x, p_y \rightarrow 0}H^2_{ttxyxy} (\omega, \vec p) = 0.
\end{equation}
Therefore we obtain 
\begin{equation}
 I_2(\omega, p = p_z) = 0 .
\end{equation}

\subsubsection*{Fourier transform: $\textbf{I}_3$}
\begin{eqnarray}
I_3(s)&=&-2\frac{d(d-2)a-(d-2)b-2c}{d(d+2)} H^2_{xyxy\alpha\beta}(s) 
\langle T_{\alpha \beta}(0) \rangle, \nonumber\\
\hat I_3(\omega, \vec p)&=&-2\frac{d(d-2)a-(d-2)b-2c}{d(d+2)} H^2_{xyxy\alpha\beta}(\omega, \vec p) 
\langle T_{\alpha \beta}(0) \rangle, \nonumber\\
H^2_{xyxyxx}(s)&=&H^2_{xyxyyy}(s)=(2-\frac{4}{d}) \partial_x \partial_y \frac{xy}{(t^2+r^2)^\frac{d}{2}},\nonumber\\
H^2_{xyxyzz}(s)+ \sum_{i = 4}^{d-1}
H^2_{xyxyii}(s)&=&-\frac{4(d-3)}{d} \partial_x \partial_y \frac{xy}{(t^2+r^2)^\frac{d}{2}}.
\end{eqnarray}
Using these inputs and the result for the Fourier transform we obtain 
\begin{eqnarray}
\sum_{ i =1}^{d-1} H^2_{xyxyii}(\omega, \vec p)
&=&\frac{ 16 \pi ^{d/2} (p_x p_y)^2  }{ d \Gamma \left(\frac{d}{2}\right) 
\left( \omega^2+ \vec p^{\, 2}\right)^2}. 
\end{eqnarray}
Thus taking the limit 
we obtain 
\begin{eqnarray}
\lim_{p_x, p_y\rightarrow 0} \sum_{ i =1}^{d-1} H^2_{xyxyii}(\omega, \vec p)
=0.
\end{eqnarray}
Similarly we have 
\begin{equation}
\lim_{p_x, p_y\rightarrow 0}  H^2_{xyxytt} (\omega, \vec p) = 0.
\end{equation}
Therefore  we conclude that the contribution of $\hat I_3$ vanishes. 
\begin{equation}
\hat  I_3(\omega, p =p_z)=0.
 \end{equation}

\subsubsection*{Fourier transform $\textbf{I}_4$}
\begin{eqnarray}
I_4(s)&=&\frac{2da+2b-c}{d(d-2)} H^3_{\alpha\beta xyxy}(s),\nonumber\\
\hat I_4(\omega, p)&=&\frac{2da+2b-c}{d(d-2)} H^3_{\alpha\beta xyxy}(\omega, p),\nonumber\\
\sum_{i = 1}^{d-1}
H^3_{ii xyxy}(\omega, p= p_z ) 
&=&\frac{4 \pi ^{d/2} \left(d \omega ^2-p ^2-\omega ^2\right)}{d \Gamma \left(\frac{d}{2}-1\right) 
\left(p ^2+\omega ^2\right)} . 
\end{eqnarray}
Similarly we have 
\begin{eqnarray} 
H^3_{tt xyxy}(\omega, p)&=& -
\frac{4 \pi ^{d/2} \left( d \omega ^2- p ^2- \omega ^2\right)}{d \Gamma \left(\frac{d}{2}-1\right)
 \left(p ^2+\omega ^2\right)}.
 \end{eqnarray}
 Therefore combining these tensor structures along with the 
 expectation value of the stress tensor we obtain 
 \begin{equation}
\hat I_4(\omega, p =p_z ) = \frac{(2 a d+2 b-c) }{ d( d-2) }
\frac{4 \pi ^{d/2} \left(d \omega ^2-p ^2-\omega ^2\right)}{d \Gamma \left(\frac{d}{2}-1\right) 
\left(p ^2+\omega ^2\right)} ( P -\epsilon_E). 
\end{equation}

\subsection*{Fourier transform: $\textbf{I}_5$}
\begin{eqnarray}
I_5(s)&=&-\frac{2(d-2)a-b-c}{d(d-2)}H^4_{\alpha\beta xyxy}(s)\langle T_{\alpha\beta}\rangle, 
\nonumber\\
\hat I_5(\omega, p)&=&-\frac{2(d-2)a-b-c}{d(d-2)}H^4_{\alpha\beta xyxy}(\omega, p)
\langle T_{\alpha\beta} \rangle. 
\end{eqnarray}
Now the tensor structure $H^4_{\alpha\beta xyxy}$ is given by 
\begin{eqnarray}
H^4_{\alpha\beta xyxy}(s)&=&\left[ (2\delta_{\alpha x}\delta_{\beta y}-\frac{2}{d}\delta_{\alpha \beta})\partial_y^2 +(2\delta_{\alpha y}\delta_{\beta y}-\frac{2}{d}\delta_{\alpha \beta})\partial_x^2\right]
 \frac{1}{(t^2+r^2)^\frac{d-2}{2}}.
 \end{eqnarray}
 Due to the presence of the external derivatives $\partial_x^2, \partial_y^2$, 
 the Fourier transform of this tensor structure will have these derivatives
 replaced by $p_x^2, p_y^2$ respectively. 
 Thus in the limit $p_x, p_y \rightarrow 0$, the Fourier transform vanishes and 
 we obtain
 \begin{eqnarray}
H^4_{iixyxy}(\omega, p=p_z ) = H^4_{ttxyxy}(\omega, p= p_z ) =0.
\end{eqnarray}
Therefore we conclude
\begin{eqnarray}
\hat I_5(\omega, p= p_z )&=&0. 
\end{eqnarray}
\subsubsection*{Fourier transform: $\textbf{I}_6$}
\begin{eqnarray}
I_6(s)&=&-\frac{2((d-2)a-c)}{d(d-2)}2H^3_{xyxy\alpha\beta}(s) \langle T_{\alpha \beta}\rangle ,\nonumber\\
H^3_{xyxy\alpha\beta}(s)&=&(\delta_{x\alpha}\delta_{y\beta}+
\delta_{y\alpha}\delta_{x\beta})\partial_x \partial_y \frac{1}{(t^2+r^2)^\frac{d-2}{2}}.
\end{eqnarray}
Again from the tensor structure of $H^3_{xyxy\alpha\beta}$, it  is clear 
that its Fourier transform will be proportional to $p_x p_y$. Therefore 
in the limit $p_x, p_y\rightarrow 0$, it vanishes. 
Thus we have 
\begin{eqnarray}
\nonumber\\
H^3_{xyxyii}(\omega, p= p_z)&=&H^3_{xyxytt}( \omega, p = p_z )=0.\nonumber\\
\end{eqnarray}.
This implies 
\begin{eqnarray}
\hat  I_6(\omega, p = p_z )&=&0.
\end{eqnarray}
\subsection*{Fourier transform: $\textbf{I}_7$}

\begin{eqnarray}
I_7(s)&=& 2\frac{((d-2)(2a+b)-dc)}{d(d^2-4)}(H^4_{xyxy\alpha\beta})(s)\langle T_{\alpha\beta} \rangle,
\nonumber\\
H^4_{xyxy\alpha\beta}(s)&=&(2\delta_{y\alpha}\partial_y\partial_\beta +
2\delta_{x\alpha}\partial_x\partial_\beta-\frac{2}{d}\delta_{\alpha\beta}(\partial_y^2+\partial_x^2)) \frac{1}{(t^2+r^2)^\frac{d-2}{2}}.\nonumber\\
\end{eqnarray}
It is clear from the tensor structure of $H^4_{xyxy\alpha\beta}$, that terms in the Fourier transform 
will always be proportional to  $p_x$ or $p_y$. 
Therfore we have 
\begin{eqnarray}
H^4_{xyxyii}(\omega, p = p_z )= 
H^4_{xyxytt}(\omega, p = p_z ) =0.
\end{eqnarray}
This allows us to conclude that the contribution
\begin{eqnarray}
\hat I_7(\omega, p = p_z )&=&0.
\end{eqnarray}
\subsubsection*{Fourier transform: $\textbf{I}_8$}

The Fourier transform of the contact term  $I_8$ leads to a constant in momentum. 
This is given by 
\begin{eqnarray}
I_8(s)&=& C h^5_{xyxy\alpha\beta} S_d \delta^d(s),\nonumber\\
h^5_{xyxy\alpha\beta}&=&2\delta_{y\alpha}\delta_{y\beta}+2\delta_{x\alpha}\delta_{x\beta}-\frac{4}{d}\delta_{\alpha\beta}.\nonumber\\
\end{eqnarray}
Performing this Fourier transform and also including the expectation value of the
stress tensor we obtain 
\begin{eqnarray}
\hat I_8(\omega, p)&=&
\frac{(d-2)(2a+b)-dc}{d^2(d+2)}\frac{8 \pi^{d/2}}{\Gamma[d/2]} ( P- \epsilon_E). 
\end{eqnarray}

\subsubsection*{Summing up the contributions}

Let us first sum up the contributions $\hat I_1, \cdots \hat I_7$, that is all the terms excluding the 
contribution from the contact term $\hat I_8$.  
This leads to 
\begin{equation} \label{hofmalcoe}
 \sum_{i = 1}^7 \hat I ( \omega, p = p_z)  = \frac{P \left(p^2-(d-1) \omega ^2\right) 
(a (d (d+4)-4)+d (2 b-c))}{2 \left(p ^2+\omega ^2\right) \left(-a \left(d^2+d-6\right)+2 b+c d+c\right)}.
\end{equation}
Here we have replaced the Euclidean energy  $\epsilon_E = - ( d-1) P$. 
Now note that this can be written as 
\begin{equation}\
 \sum_{i = 1}^7 \hat I ( \omega, p ) = 2 a_{T, 0}  \frac{ (p^2-(d-1) \omega ^2}{p^2 + \omega^2} P, 
\end{equation}
where $c_{T, 0}$ is the Hofman-Maldacena coefficient 
in the scalar channel given in equation (2.16) of \cite{Komargodski:2016gci}. 

Though our analysis here has been for $d>3$, we can carry out the same steps for
 $d=3$, with momentum say in the $y$ direction. 
Now the values of  Fourier transforms $\hat I_1, \hat I_2, \hat I_4, \hat I_5, \hat I_7$ are 
non-zero while $\hat I_3, \hat I_6$ vanish
However on summing up their contributions
we obtain the result
\begin{equation}
  \sum_{i = 1}^7 \hat I_i ( \omega, 0  ) =
 \frac{
 17 a+6 b-3 c }{6 a-2 (b+2 c) } P. 
\end{equation}
This is indeed the same result as that obtained by taking the $p\rightarrow 0$ limit of 
the expression in  (\ref{hofmalcoe}) with $d=3$.

\subsection{ The vector channel} \label{fouriertransformvector}

We now study the Fourier transform in the vector channel. 
We examine   the  various tensor structures corresponding to the 
$T_{xt} T_{xz}$ OPE with momentum along the $p_z$ directions. 
Such a kinematic configuration is possible for all $d\geq 3$ dimensions.

\subsubsection*{Fourier transform: $\textbf{I}_1$}
\begin{eqnarray}
I_1(s)&=&\frac{(d-2)(4a+2b-c)}{(d+2)}H^1_{\alpha \beta xtxz}(s) 
\langle T_{\alpha \beta}(0) \rangle, \\
H^1_{\alpha \beta xtxz}(s)&=&(\partial_\alpha \partial_\beta - \frac{1}{d}
\delta_{\alpha \beta} \partial^2)\left(\frac{x^2tz}
{(t^2+ \vec r^{\, 2})^\frac{d+2}{2}}\right).\nonumber
\end{eqnarray}
Using the result in (\ref{inttype7}) we obtain for the Fourier transform
\begin{eqnarray}\nonumber
\sum_{i =1}^{d-1} H^1_{iixtxz}(\omega, \vec p)&=&\left(- \vec p^{\, 2}  
+\frac{(d-1)}{d}(\omega^2 +\vec p^{\, 2} )\right)
\left(-\frac{2 \pi ^{d/2} p_z \omega  \left( \vec p^{\, 2} + \omega^2  - 4 p_x^2 \right)}
{\left( \vec p^{\, 2} +\omega^2\right)^3 \Gamma\left[1+\frac{d}{2}\right]}\right).\\
\end{eqnarray}
Now taking the limit of all momenta to vanish except $p_z$ we obtain 
\begin{eqnarray}
\sum_{i=1}^{d-1} H^1_{iixtxz}(\omega,  p=p_z) 
&=&-\frac{2 \left(\pi ^{d/2} p_z \omega  \left(d \omega ^2-p_z^2-\omega ^2\right)\right)}
{d \Gamma \left(\frac{d}{2}+1\right) \left(\omega^2+p_z^2\right)^2}.
\end{eqnarray}
Similarly for  the time component we obtain 
\begin{eqnarray}
H^1_{ttxtxz}(\omega , p =p_z)&=&
\frac{2 \left(\pi ^{d/2} p_z \omega  \left(d \omega ^2-p_z^2-\omega ^2\right)\right)}
{d \Gamma \left(\frac{d}{2}+1\right) \left(\omega^2+p_z^2\right)^2}.
\end{eqnarray}
Using these results and substituting the values of the expectation value of the stress tensor we obtain 
\begin{eqnarray}
\hat I_1(\omega, p_z)
&=&\frac{(d-2) (4 a+2 b-c)}{d+2} \frac{2 \left(\pi ^{d/2} p_z \omega  \left(d \omega ^2-p_z^2-\omega ^2\right)\right)}
{d \Gamma \left(\frac{d}{2}+1\right) \left(\omega^2+p_z^2\right)^2}( - P + \epsilon_E). 
\end{eqnarray}

\subsubsection*{Fourier transform: $\textbf{I}_2$}
\begin{eqnarray}
I_2(s)&=& \frac{1}{d}(da+b-c) H^2_{\alpha \beta xtxz}(s)\langle T_{\alpha \beta}(0) \rangle, \nonumber\\
H^2_{\alpha \beta xtxz}(s)&=&(\partial_z\partial_t)\frac{1}{(t^2+r^2)^\frac{d-2}{2}}(\frac{s_\alpha s_\beta}{t^2+r^2}-\frac{1}{d} \delta_{\alpha \beta}).\nonumber\\
\end{eqnarray}
We can use the result in (\ref{inttype2})  to carry out the Fourier transform. 
When all momenta except $p_z$ is set to zero we obtain 
\begin{eqnarray}
\sum_{i =1}^{d-1} H^2_{iixtxz}(\omega, p = p_z )&=&
\frac{2 \pi ^{d/2} p_z \omega  \left(p_z^2-(-1+d) \omega ^2\right)}
{\left(p_z^2+\omega ^2\right)^2 \Gamma\left[1+\frac{d}{2}\right]} . 
\end{eqnarray}
Similarly  for the time component we obtain 
\begin{eqnarray}
H^2_{ttxtxz}(\omega, p= p_z)&=& -\frac{2 \pi ^{d/2} p_z \omega 
\left(p_z^2-(-1+d) \omega ^2\right)}
{\left(p_z^2+\omega ^2\right)^2 \Gamma\left[1+\frac{d}{2}\right]}.
\end{eqnarray}
We can now substitute these expressions along with the expectation value of the 
stress tensor to obtain the Fourier transform
\begin{eqnarray}
\hat I_2(\omega, p =p_z)&=&\frac{(da+b-c)}{d}
\left(\frac{2 \pi ^{d/2} p_z \omega  \left(p_z^2-(-1+d) \omega ^2\right)}{\left(p_z^2+\omega ^2\right)^2 
\Gamma\left[1+\frac{d}{2}\right]}\right) ( P -\epsilon_E) .
\end{eqnarray}

\subsubsection*{Fourier transform: $\textbf{I}_3$}
\begin{eqnarray}
I_3(s)&=&-\frac{d(d-2)a-(d-2)b-2c}{d(d+2)} (H^2_{xtxz\alpha\beta}(s) +H^2_{xzxt\alpha\beta}(s))\langle T_{\alpha \beta}(0)  \rangle. 
\end{eqnarray}
Now the tensor structures in $I_3$ are given by
\begin{eqnarray}
H^2_{xtxz\alpha\beta}(s)&=&(\delta_{x\alpha}\partial_z \partial_\beta +\delta_{x\beta}\partial_z \partial_\alpha + \delta_{z\beta}\partial_x \partial_\alpha + \delta_{z\alpha}\partial_x \partial_\beta -\frac{4}{d} \delta_{\alpha\beta}\partial_x\partial_z) \frac{xt}{(t^2+r^2)^{\frac{d}{2}}},\nonumber\\
H^2_{xzxt\alpha\beta}(s)&=&(\delta_{x\alpha}\partial_t \partial_\beta + \delta_{x\beta}\partial_t \partial_\alpha + \delta_{t\alpha}\partial_x \partial_\beta + \delta_{t\beta}\partial_x \partial_\alpha-\frac{4}{d} \delta_{\alpha\beta}\partial_x\partial_t) \frac{xz}{(t^2+r^2)^{\frac{d}{2}}}.\nonumber\\
\end{eqnarray}
We can use (\ref{inttype6}) to Fourier transfrom these tensor structures. 
However it is easy to see that all of the terms  which 
occur in the Fourier transform are proportional to momenta
orthogonal to $p_z$. Therefore they all vanish when only $p_z$ is turned on
which implies 
\begin{equation}
H^2_{xtxz\alpha\alpha}(\omega, p=p_z) =  
 H^2_{xzxt\alpha\beta}( \omega, p = p_z) =0.
 \end{equation}
 Therefore we obtain 
 \begin{equation}
 \hat I_3(\omega, p = p_z) =0.
\end{equation}
\subsubsection*{Fourier transform: $\textbf{I}_4$}
\begin{eqnarray}
I_4(s)&=&\frac{2da+2b-c}{d} H^3_{\alpha \beta xtxz}(s) \langle T_{\alpha\beta} \rangle, 
\nonumber\\
H^3_{\alpha \beta xtxz}(s)&=& h^3_{xtxz}(\partial_\alpha \partial_\beta-\frac{1}{d}\delta_{\alpha\beta} \partial^2) \frac{1}{(t^2+r^2)^{\frac{d-2}{2}}}.
\end{eqnarray}
Now $h^3$ is defined as
\begin{equation}\label{defh3}
h^3_{\mu\nu\rho\sigma} = \delta_{\mu\rho} \delta_{\nu\sigma} +
 \delta_{\mu\sigma}\delta_{\nu\rho} - \frac{2}{d} \delta_{\mu\nu} \delta{\rho\sigma}.
 \end{equation}
 Therefore the  component $h^3_{xtxz} =0$, 
 which leads to to conclude
 \begin{equation}
 \hat I_4(\omega, p_z )=0. 
\end{equation}
\subsubsection*{Fourier transform: $\textbf{I}_5$}
\begin{eqnarray}
I_5(s)&=&-\frac{2(d-2)a-b-c}{d(d-2)}H^4_{\alpha\beta xtxz}(s)\langle T_{\alpha\beta}\rangle,
\nonumber\\
H^4_{\alpha\beta xtxz}(s)&=&((2\delta_{\alpha x}\delta_{\beta x}-\frac{2}{d}\delta_{\alpha \beta})\partial_z\partial_t) \frac{1}{(t^2+r^2)^{\frac{d-2}{2}}}.
\end{eqnarray}
The Fourier transform can be done using the result in (\ref{inttype4})
\begin{eqnarray}
\sum_{i=1}^{d-1}
H^4_{iixtxz}(\omega, p_z )&=&\int d^dx e^{-i\omega t- i p_z z}(\frac{2}{d})
\partial_z\partial_t \frac{1}{(t^2+r^2)^{\frac{d-2}{2}}},\nonumber\\
&=& -\frac{8 \left(\pi ^{d/2} p_z \omega \right)}{d \left(p_z^2+\omega^2\right) \Gamma\left[-1+\frac{d}{2}\right]}.
\end{eqnarray}
Similarly we have 
\begin{eqnarray}
H^4_{ttxtxz}(\omega, p_z)&=&\frac{8 \pi ^{d/2} p_z \omega }{d \left(p_z^2+\omega^2\right) \Gamma\left[-1+\frac{d}{2}\right]}.
\end{eqnarray}
Using these results along with the expectation values of the stress tensor 
we obtain 
\begin{eqnarray}
\hat I_5(\omega, p_z)&=&\frac{-(2(d-2)a-b-c)}{d(d-2)} 
\left(\frac{8 \left(\pi ^{d/2} p_z \omega \right)}{d \left(p_z^2+\omega^2\right) \Gamma\left[-1+\frac{d}{2}\right]}\right)( -P+ \epsilon_E)  .  \nonumber \\
\end{eqnarray}
\subsubsection*{Fourier transform: $\textbf{I}_6$}
\begin{eqnarray}
I_6(s)&=&-\frac{2((d-2)a-c)}{d(d-2)}(H^3_{xtxz\alpha\beta}(s)+H^3_{xzxt\alpha\beta}(s)
\langle T_{\alpha \beta} \rangle, \nonumber\\
H^3_{xtxz\alpha\beta}(s)&=&h^3_{xz\alpha\beta}\partial_x\partial_t\frac{1}{(t^2+r^2)^{\frac{d-2}{2}}},\nonumber\\
H^3_{xzxt\alpha\beta}(s)&=&h^3_{xt\alpha\beta}\partial_x\partial_z\frac{1}{(t^2+r^2)^{\frac{d-2}{2}}}.
\end{eqnarray}
From the definition of $h^3$ in (\ref{defh3}) it is clear for that for $\alpha=\beta$, the above 
components of $h^3$ vanish. 
Thus we obtain 
\begin{eqnarray}
\hat I_6(\omega, p_z )&=&0. 
\end{eqnarray}
\subsubsection*{Fourier transform: $\textbf{I}_7$}
\begin{eqnarray}
I_7(s)&=& \frac{((d-2)(2a+b)-dc)}{d(d^2-4)}(H^4_{xtxz\alpha\beta}+H^4_{xzxt\alpha\beta})(s)\langle T_{\alpha\beta} \rangle.
\end{eqnarray}
The tensor structures involved in $I_7$ are given by 
\begin{eqnarray}
H^4_{xzxt\alpha\beta}(s)&=&(\delta_{z\alpha}\partial_t\partial_\beta +\delta_{z\beta}\partial_t\partial_\alpha -\frac{2}{d}\delta_{\alpha\beta}(\partial_z\partial_t)) \frac{1}{(t^2+r^2)^{\frac{d-2}{2}}},\nonumber\\
H^4_{xtxz\alpha\beta}(s)&=&(\delta_{t\alpha}\partial_z\partial_\beta +\delta_{t\beta}\partial_z\partial_\alpha -\frac{2}{d}\delta_{\alpha\beta}(\partial_z\partial_t)) \frac{1}{(t^2+r^2)^{\frac{d-2}{2}}}.
\end{eqnarray}
Now we can Fourier transform these tensors using the result (\ref{inttype4}) 
\begin{eqnarray}
\sum_{i=1}^{d-1}H^4_{xtxzii}(\omega, p_z)
&=& \frac{8 (-1+d) \pi ^{d/2} p_z \omega }{d \left(p_z^2+\omega^2\right) \Gamma\left[\frac{1}{2} (-2+d)\right]} , \nonumber\\
H^4_{xtxztt}(\omega, p_z)&=& -\frac{8 \left((-1+d) \pi ^{d/2} p_z \omega \right)}{d \left(p_z^2+\omega^2\right) \Gamma\left[\frac{1}{2} (-2+d)\right]},\nonumber\\
\sum_{i =1}^{d-1}
H^4_{xzxtii}(\omega, p)&=& -\frac{8 \left(\pi ^{d/2} p_z \omega \right)}{d \left(p_z^2+\omega^2\right) \Gamma\left[\frac{1}{2} (-2+d)\right]}, \nonumber\\
H^4_{xzxttt}(\omega, p)&=& \frac{8 \pi ^{d/2} p_z \omega }{d \left(p_z^2+\omega^2\right) \Gamma\left[\frac{1}{2} (-2+d)\right]}.
\end{eqnarray}
Now using these results for the Fourier transforms as well as the
expectation value of the stress tensor we obtain 
\begin{eqnarray}
\hat I_7(\omega, p_z)&=&
\frac{ (d-2)(2a+b)-dc }{d(d^2-4)} 
\left(\frac{8 (-2+d) \pi ^{d/2} p_z \omega }{d \left(p_z^2+\omega^2\right) \Gamma\left[\frac{1}{2} (-2+d)\right]}  \right) ( P - \epsilon_E) . \nonumber \\
\end{eqnarray}

\subsubsection*{Fourier transform: $\textbf{I}_8$}
Finally the contribution from the contact term reduces to 
\begin{eqnarray}
I_8(s)&=& C h^5_{xtxz\alpha\beta} S_3 \delta^3(s),\nonumber\\
h^5_{xtxz\alpha\beta}&=&\delta_{t\alpha}\delta_{z\beta}-\frac{4}{d}\delta_{\alpha\beta}h^3_{xtxz}.
\end{eqnarray}
Using the definition of $h^3$ in (\ref{defh3}) we see that $h^5_{xtxz\alpha\alpha}$ vanishes. 
Thus we obtain 
\begin{equation}
\hat I_8(\omega, p_z )=0 
\end{equation}

\subsubsection*{Summing up the contributions}

Summing up all the contributions we obtain 
\begin{eqnarray}
\hat A_{x\tau xz} (\omega, p_z) &=& \sum_{i =1}^8 \hat I_i ( \omega, p_z) , \\ \nonumber
&=& -\frac{p_z \omega}{(p_z^2+\omega^2)} G_2(\omega, p_z),
\end{eqnarray}
where 
\begin{eqnarray}
G_2(\omega, p_z) = - \delta G_R(0)  +   a_{T, 1}\frac{ 8 d p_z^2}{\omega^2 + p_z^2} P.
\end{eqnarray}
$\delta G_R(0)$ is the RHS of the sum rule defined in (\ref{sumabc}) 
and $a_{T, 1}$ is the Hofman-Maldacena coefficient in the vector channel defined  as
\begin{equation}
a_{T, 1} = \frac{1}{8} \frac{ b (2-3 d)+2 c d-a (-8+d (6+d))  }{- (2 b+c+c d)+ a \left(-6+d+d^2\right) }.
\end{equation}

\subsection{The sound channel}\label{fouriertransformscalar}

In this part of the appendix we perform the Fourier transform of the 
OPE coefficient in the sound channel by the considering the
$T_{tt} T_{tt}$ OPE . We examine all the structures which occur
in the coefficients $\hat A_{tttt\alpha\beta} \langle T_{\alpha\beta}\rangle$ 
term by term and then sum them together.  We choose momentum to be along $p_z$. 

\subsubsection*{Fourier transform: $\textbf{I}_1$}
\begin{eqnarray}
I_1(s)&=&\frac{(d-2)(4a+2b-c)}{(d+2)}H^1_{\alpha \beta tttt}(s) 
\langle T_{\alpha \beta}(0) \rangle, \nonumber\\
H^1_{\alpha \beta tttt}(s)&=&(\partial_\alpha \partial_\beta - \frac{1}{d} \delta_{\alpha \beta} \partial^2)\left[ \left( \frac{t^2}{ t^2 + \vec r^{\, 2} } - \frac{1}{d} \right)^2 
 \frac{1}{(  t^2 + \vec r^2)^{ \frac{d-2}{2} }}   \right]. 
 \end{eqnarray}
 We can use integrals of the type in (\ref{inttype7}) to obtain 
 \begin{eqnarray}
 \sum_{i =1}^{d-1} H^1_{iitttt}(\omega, p_z)
=-\frac{4 \pi ^{d/2} \left[p_z^2-(-1+d) \omega ^2\right] \left[(-1+d) p_z^4-2 (1+d) p_z^2 \omega ^2+(-1+d) \omega ^4\right]}{d^3 \left(p_z^2+\omega ^2\right)^3 \Gamma\left[\frac{d}{2}\right]}.
\nonumber \\
\end{eqnarray}
Similarly  one finds 
\begin{eqnarray}
H^1_{ttttt}(\omega , p_z )&=& - \sum_{i =1}^{d-1}H^1_{iitttt}(\omega, p_z).
\end{eqnarray}

Using these results together with the expectation value of the stress tensor we obtain
\begin{eqnarray}
&& \hat I_1(\omega, p_z )=\frac{(d-2)}{d+2}(4a+2b-c) \times \\ \nonumber
&&\left(  \frac{4 \pi ^{d/2} \left[p_z^2-(-1+d) \omega ^2\right] \left[(-1+d) p_z^4-2 (1+d) p_z^2 \omega ^2+(-1+d) \omega ^4\right]}{d^3 \left(p_z^2+\omega ^2\right)^3 \Gamma\left[\frac{d}{2}\right]}
\right) ( - P + \epsilon_E). 
\end{eqnarray}

\subsubsection*{Fourier transform: $\textbf{I}_2$}
\begin{eqnarray}
I_2(s)&=& \frac{1}{d}(da+b-c) H^2_{\alpha \beta tttt}(s)\langle T_{\alpha \beta}(0) \rangle,
\nonumber\\
H^2_{\alpha \beta tttt}(s)&=&((4-\frac{8}{d})\partial_t^2+\frac{4}{d^2}\partial^2)\frac{1}{(t^2+r^2)^\frac{d-2}{2}}(\frac{s_\alpha s_\beta}{t^2+r^2}-\frac{1}{d} \delta_{\alpha \beta}).
\end{eqnarray}
Now using the result in (\ref{inttype2}) to perform the Fourier transform 
we obtain
\begin{eqnarray}
\sum_{i =1}^{d-1}
H^2_{iitttt}(\omega, p = p_z)
&=&\frac{16 \pi ^{d/2} \left(p_z^2-(-1+d) \omega ^2\right) \left(p_z^2+(-1+d)^2 \omega ^2\right)}{d^3 \left(p_z^2+\omega ^2\right)^2 \Gamma\left[\frac{d}{2}\right]}.
\end{eqnarray}
Similarly we obtain
\begin{eqnarray}
H^2_{tttttt}(\omega, p=p_z)&=&- \sum_{i =1}^{d-1}H^2_{iitttt}(\omega, p = p_z).
\end{eqnarray}
Then substituting these Fourier transforms and taking the expectation 
value of the stress tensor
we obtain
\begin{eqnarray}
\hat I_2(\omega, p = p_z) 
 &=&\frac{(da+b-c)}{d}
 \left(\frac{16 \pi ^{d/2} \left(p_z^2-(-1+d) \omega ^2\right) \left(p_z^2+(-1+d)^2 \omega ^2\right)}{d^3 \left(p_z^2+\omega ^2\right)^2 \Gamma\left[\frac{d}{2}\right]}\right)(P- \epsilon_E). \nonumber
 \\
\end{eqnarray}

\subsubsection*{Fourier transform: $\textbf{I}_3$}
\begin{eqnarray}
I_3(s)&=&-2\frac{d(d-2)a-(d-2)b-2c}{d(d+2)} H^2_{tttt\alpha\beta}(s) 
\langle T_{\alpha \beta}(0) \rangle,\\ \nonumber
H^2_{tttt\alpha\beta}(s)&=&(2\delta_{t\alpha}\partial_t \partial_\beta +2\delta_{t\beta}\partial_t \partial_\alpha -\frac{4}{d} \partial_\alpha \partial_\beta -\frac{4}{d} \delta_{\alpha\beta}\partial_t^2 +\frac{4}{d^2} \partial^2 ) \frac{1}{(t^2+r^2)^{\frac{d-2}{2}}}(\frac{t^2}{t^2+r^2}-\frac{1}{d}).
\end{eqnarray}
Using the result for the Fourier transform in (\ref{inttype2}) 
we obtain
\begin{eqnarray}
\sum_{i =1}^{d-1}
H^2_{ttttii}(\omega, p_z)&=&\frac{8 \pi ^{d/2} \left(p_z^2+\omega ^2-d \omega ^2\right) \left(p_z^2+\omega ^2-2 d \omega ^2+d^2 \omega ^2\right)}{d^2 \left(p_z^2+\omega ^2\right)^2 \Gamma\left[1+\frac{d}{2}\right]}, \nonumber\\
H^2_{tttttt}(\omega, p_z)&=& -\sum_{i =1}^{d-1}
H^2_{ttttii}(\omega, p_z). 
\end{eqnarray}
Substituting these Fourier transforms and taking the expectation values of the
stress tensor we obtain 
\begin{eqnarray}
& & \hat I_3(\omega, p_z)=-2\frac{(d(d-2)a-(d-2)b-2c)}{d(d+2)} \times  \\ \nonumber
 && \qquad 
 \left( \frac{8 \left(\pi ^{d/2} \left(p_z^2+\omega ^2-d \omega ^2\right) \left(p_z^2+\omega ^2-2 d \omega ^2+d^2 \omega ^2\right)\right)}{d^2 \left(p_z^2+\omega ^2\right)^2 \Gamma\left[1+\frac{d}{2}\right]}\right)( P - \epsilon_E) .
\end{eqnarray}
\subsubsection*{Fourier transform: $\textbf{I}_4$}
\begin{eqnarray}
I_4(s)&=&\frac{2da+2b-c}{d} H^3_{\alpha \beta tttt}(s)\langle T_{\alpha\beta} \rangle,
\nonumber\\
H^3_{\alpha \beta tttt}(s)&=& h^3_{tttt}(\partial_\alpha \partial_\beta-\frac{1}{d}\delta_{\alpha\beta} \partial^2) \frac{1}{(t^2+r^2)^{\frac{d-2}{2}}}.
\end{eqnarray}
Using (\ref{inttype1}) for the Fourier transform we obtain 
\begin{eqnarray}
\sum_{i =1}^{d-1}
H^3_{iitttt}( \omega, p_z )&=&\frac{8 (-1+d) \pi ^{d/2} \left(-p_z^2+(-1+d) \omega ^2\right)}{d^2 \left(p_z^2+\omega ^2\right) \Gamma\left[-1+\frac{d}{2}\right]},\nonumber\\
H^3_{tttttt}(  \omega, p_z ) &=&   - 
\sum_{i =1}^{d-1}
H^3_{iitttt}( \omega, p_z ).
\end{eqnarray}
Finally substituting the expectation values of the stress tensor we obtain 
\begin{eqnarray}
\hat I_4(\omega, p_z)&=&
\frac{(2da+2b-c)}{d(d-2)}
\left(\frac{8 (-1+d) \pi ^{d/2} \left(-p_z^2+(-1+d) \omega ^2\right)}{d^2 \left(p_z^2+\omega ^2\right) \Gamma\left[-1+\frac{d}{2}\right]}\right)( P  - \epsilon_E). \nonumber \\
\end{eqnarray}
\subsubsection*{Fourier transform: $\textbf{I}_5$}
\begin{eqnarray}
I_5(s)&=&-\frac{2(d-2)a-b-c}{d(d-2)}H^4_{\alpha\beta tttt}(s)\langle T_{\alpha\beta},
\rangle \\
\nonumber
H^4_{\alpha\beta tttt}(s)&=&(4h^3_{\alpha\beta tt}\partial_t^2 - \frac{8}{d}h^3_{\alpha\beta\lambda t}\partial_\lambda \partial_t + \frac{8}{d^2}(\partial_\alpha\partial_\beta -\frac{1}{d}\delta_{\alpha\beta}\partial^2)) \frac{1}{(t^2+r^2)^{\frac{d-2}{2}}}.
\end{eqnarray}
We can Fourier transform using (\ref{inttype1}) and we obtain 
\begin{eqnarray}
\sum_{i=1}^{d-1}
H^4_{iitttt}(\omega, p_z)&=&-\frac{32 \left(\pi ^{d/2} \left(p_z^2-(-1+d)^3 \omega ^2\right)\right)}{d^3 \left(p_z^2+\omega ^2\right) \Gamma\left[-1+\frac{d}{2}\right]},\nonumber\\
H^4_{tttttt}(\omega, p_z)&=& - \sum_{i=1}^{d-1}
H^4_{iitttt}(\omega, p_z).
\end{eqnarray}
Using these results for the transforms along with the expectation 
values of the stress tensor we obtain
\begin{eqnarray}
\hat I_5(\omega, p)&=&\frac{-(2(d-2)a-b-c)}{d(d-2)}
\left( \frac{32 \pi ^{d/2} \left(p_z^2-(-1+d)^3 \omega ^2\right) }{d^3 \left(p_z^2+\omega ^2\right) \Gamma\left[-1+\frac{d}{2}\right]}\right)( -P + \epsilon_E) . \nonumber \\
\end{eqnarray} 
\subsubsection*{Fourier transform: $\textbf{I}_6$}

\begin{eqnarray}
I_6(s)&=&-\frac{4((d-2)a-c)}{d(d-2)}H^3_{tttt\alpha\beta}(s)\langle T_{\alpha\beta} \rangle,
\nonumber\\
H^3_{tttt\alpha\beta}(s)&=&h^3_{tt\alpha\beta}(\partial_t^2-\frac{1}{d}\partial^2)\frac{1}{(t^2+r^2)^{\frac{d-2}{2}}}.
\end{eqnarray}
The Fourier transform is done by using (\ref{inttype1}).
\begin{eqnarray}
\sum_{i =1}^{d-1}H^3_{ttttii}(\omega, p_z)
&=&-\frac{8 (-1+d) \pi ^{d/2} \left(p_z^2-(-1+d) \omega ^2\right)}{d^2 \left(p_z^2+\omega ^2\right) \Gamma\left[-1+\frac{d}{2}\right]},\nonumber\\
H^3_{tttttt}(\omega, p_z )&=& -\sum_{i =1}^{d-1}H^3_{ttttii}(\omega, p_z).
\end{eqnarray}
Substituting the Fourier transform along with the expectation value of the
stress tensor we obtain
\begin{eqnarray}
 \hat I_6(\omega, p_z)&=&-4\frac{((d-2)a-c)}{d(d-2)} 
 \left(\frac{8  (-1+d) \pi ^{d/2} \left(p_z^2-(-1+d) \omega ^2\right)
 }{d^2 \left(p_z^2+\omega ^2\right) \Gamma\left[-1+\frac{d}{2}\right]}\right)( -P + \epsilon_E). 
 \nonumber \\
\end{eqnarray}
\subsubsection*{Fourier transform $\textbf{I}_7$}
\begin{eqnarray}
\hat I_7(\omega, p_z)&=& \frac{2((d-2)(2a+b)-dc)}{d(d^2-4)}(H^4_{tttt\alpha\beta})(s)T_{\alpha\beta},\nonumber\\
H^4_{tttt\alpha\beta}(s)&=&\left(2h^3_{ttt\alpha}\partial_t\partial_\beta +2h^3_{ttt\beta}\partial_t\partial_\alpha -\frac{2}{d}(h^3_{tt\lambda\alpha}\partial_\lambda\partial_\beta + h^3_{tt\lambda\beta}\partial_\lambda\partial_\alpha) -\frac{2}{d}\delta_{\alpha\beta}(2h^3_{tt\lambda t}\partial_\lambda\partial_t)\right.\nonumber\\
&&\left. -\frac{8}{d^2} \delta_{\alpha \beta} (\partial_t^2-\frac{1}{d}\partial^2)\right) \frac{1}{(t^2+r^2)^{\frac{d-2}{2}}}.
\end{eqnarray}
The Fourier transform is performed using (\ref{inttype1}) 
\begin{eqnarray}
\sum_{i=1}^{d-1}
H^4_{ttttii}(\omega, p_z)&=& -\frac{32 \pi ^{d/2} \left(p_z^2-(-1+d)^3 \omega ^2\right)}{d^3 \left(p_z^2+\omega ^2\right) \Gamma\left[-1+\frac{d}{2}\right]}, \nonumber\\
H^4_{tttttt}(\omega, p_z)&=& -\sum_{i=1}^{d-1}
H^4_{ttttii}(\omega, p_z).
\end{eqnarray}
Using the results for the Fourier transform along with the expectation value
of the stress tensor we obtain
\begin{eqnarray}
\hat I_7(\omega, p_z)&=&2\frac{((d-2)(2a+b)-dc)}{d(d-2)}
\left(\frac{32 \pi ^{d/2} \left(p_z^2-(-1+d)^3 \omega ^2\right)}{d^3 \left(p_z^2+\omega ^2\right) \Gamma\left[-1+\frac{d}{2}\right]}\right)( - P  + \epsilon_E). \nonumber \\
\end{eqnarray}

\subsubsection*{Fourier transform: $\textbf{I}_8$}
Finally we Fourier transform the contact term which is given by 
\begin{eqnarray}
I_8(s)&=& (C h^5_{tttt\alpha\beta} +D(2h^3_{tt\alpha\beta})) S_d \delta^d(s)\langle 
T_{\alpha\beta} \rangle,
\nonumber\\
C&=&\frac{(d-2)(2a+b)-dc}{d(d+2)}, \qquad 
D=\frac{8 (-2 b-c (1+d)+a (-2+d) (3+d)) \pi ^{d/2}}{d^2 (2+d) \Gamma\left[\frac{d}{2}\right]}, 
\nonumber\\
h^5_{tttt\alpha\beta}&=&8\delta_{t\alpha}\delta_{t\beta} - \frac{8}{d} h^3_{tt\alpha\beta} - \frac{8}{d^2}\delta_{\alpha\beta}-\frac{4}{d}(2-\frac{2}{d})\delta_{\alpha\beta}.
\end{eqnarray}
The Fourier transform is trivial to perform and keeping track of the 
tensor structures  along with the expectation value of the stress tensor we obtain
\begin{eqnarray}
I_8(\omega, p_z)&=&\frac{2 \pi ^{d/2}}{\Gamma\left[\frac{d}{2}\right]}
\left[ 4D   \left( \frac{  1-d  }{d} \right)  + C \left(-8+ \frac{24}{d}-\frac{16}{d^2}\right) \right] ( P - \epsilon_E) 
\end{eqnarray}

\subsubsection*{Summing  up the contributions}

We now sum up all the contributions in the sound channel of the 
term $ \hat A_{tttt\alpha\beta} (\omega, p_z)  \langle T_{\alpha\beta} \rangle$ 
in the OPE. We can write 
 The sum  can be organized as 
\begin{eqnarray}
 \hat A_{tttt\alpha\beta} (\omega, p_z)  \langle T_{\alpha\beta} \rangle &=&  
 \sum_{i =1}^8 \hat I_8(\omega, p_z) , \\ \nonumber
&=& \frac{p_z^4}{(p_z^2+\omega^2)^2}G_3( \omega, p_z).
\end{eqnarray}
The function $G_3(\omega, p_z) $ admits an Laurent expansion  in $( \frac{p_z}{\omega})^2$
which is given by 
\begin{eqnarray}
G_3( \omega, p_z) &=& \left(  F_1 ( \frac{\omega}{p_z}  )^4  
+ F_2 ( \frac{\omega}{p_z}  )^2 + F_3 +   a_{T, 2}  \frac{  64(\frac{ p_z}{\omega})^2  }
{ 1 + (\frac{p_z}{\omega})^2} \right) P.
\end{eqnarray}
Here $F_1, F_2, F_3$ are ratios of linear functions of the constants $a, b, c$. 
They can be written in the variables $t_2, t_4$. 
However starting from the term $ (\frac{ p_z}{\omega})^2 $  the entire expansion 
is determined by 
$a_{T, 2}$, the Hofman-Maldacena coefficient in the tensor channel
which is given by 
\begin{equation}
a_{T, 2} = -\frac{1}{32}
\frac{ (4 a+2 b-c) (-2+d) d   }{\left(-2 b-c (1+d)+a \left(-6+d+d^2\right)\right)}.
\end{equation}

\section{Integrals}\label{integrals}

In this section we will evaluate the generic integrals that at are required to 
obtain the Fourier transform of the tensor structures that occur in 
the OPE coefficient $\hat A_{\mu\nu\rho\sigma\alpha\beta}$. 
The Fourier transforms are done with momentum turned on in 
arbitrary directions. 
To perform the transform we first 
convert the integral to polar coordinates in the spatial $d-1$  directions. 
After performing the angular integrals with integrate the radial direction 
and the time direction. We have verified that the result is independent 
of the order of the integrations. 
\subsection*{Type 1}

\begin{eqnarray}
&& \textrm{F.T}\, \left[ \partial_x^2 \frac{1}{(t^2+r^2)^\frac{d-2}{2}} \right] =  
\int d^dx \exp[-i \vec{p}\cdot \vec{r}-i \omega t] (-p_x^2) \frac{1}{(t^2+r^2)^\frac{d-2}{2}} ,\\
\nonumber
&=& (-p_x^2)\int dr dt\sqrt{\pi } r^{d-2} \Gamma \left(\frac{d}{2}-1\right) e^{-i t \omega } \left(r^2+t^2\right)^{1-\frac{d}{2}}\frac{2 \pi ^{\frac{d}{2}-1}}{\Gamma \left(\frac{d}{2}-1\right)}
\, _0\tilde{F}_1\left(;\frac{d-1}{2};-\frac{1}{4} p^2 r^2\right).
\end{eqnarray}
Here $\, _0\tilde{F}_1(; b, z)$ is the regularized hypergeometric function
defined as 
\begin{equation}
\, _0\tilde{F}_1(; b, z) = \sum_{k =0}^\infty \frac{ z^k}{ \Gamma ( b+k) k!}.
\end{equation}
Now performing the radial and the time integrations we obtain
\begin{eqnarray} \label{inttype1}
 \textrm{F.T}\, \left[ \partial_x^2 \frac{1}{(t^2+r^2)^\frac{d-2}{2}} \right] 
&=& (-p_x^2)\frac{4 \pi ^{d/2}}{\Gamma \left(\frac{d}{2}-1\right) \left( \omega^2+p^2\right)}.
\end{eqnarray}
Similarly we have the Fourier transforms
\begin{eqnarray} 
\textrm{F.T} \, 
\left[ \partial_i^2 \frac{1}{(t^2+r^2)^\frac{d-2}{2}} \right] &=& (-p_i^2)\frac{4 \pi ^{d/2}}{\Gamma \left(\frac{d}{2}-1\right) \left( \omega^2+p^2\right)} ,\\
\textrm{F.T} \, 
 \left[ \partial_t^2 \frac{1}{(t^2+r^2)^\frac{d-2}{2}} \right]
&=& (-\omega^2)\frac{4 \pi ^{d/2}}{\Gamma \left(\frac{d}{2}-1\right) \left( \omega^2+p^2\right)}.\nonumber
\end{eqnarray}           
Note here and in the rest of this appendix $p^2 = \vec p^{\, 2}$.                 
                      
\subsection*{Type 2}                      
                            
\begin{eqnarray}\label{inttype2}     
&& \textrm{F.T} \, \left[ \frac{r^2}{(t^2+r^2)^\frac{d}{2}} \right] =\int d^d x \exp[-i \vec{p}\cdot
\vec{r}-i \omega t] (\frac{r^2}{(t^2+r^2)^\frac{d}{2}}),\nonumber\\
&=&\int dr dt \sqrt{\pi } r^{d} \Gamma \left(\frac{d}{2}-1\right) e^{-i t \omega } \left(r^2+t^2\right)^{-\frac{d}{2}} \, _0\tilde{F}_1\left(;\frac{d-1}{2};-\frac{1}{4} p^2 r^2\right)(\frac{2 \pi ^{\frac{d}{2}-1}}{\Gamma \left(\frac{d}{2}-1\right)}),\nonumber\\
&=& \frac{2 \pi ^{d/2} \left((d-3) p^2  +(d-1) \omega ^2\right)}{\Gamma \left(\frac{d}{2}\right) \left(  \omega ^2+ p^2 \right)^2}.
\\ \nonumber
& & \textrm{F.T}\,  \left[ \frac{t^2}{(t^2+r^2)^\frac{d}{2}} \right]=\frac{2 \pi ^{d/2} \left(-\omega^2+
p^2 \right)}{\Gamma \left(\frac{d}{2}\right) \left( \omega^2+
p^2 \right)^2}.
\end{eqnarray}

 \subsection*{Type 3}                       
    
\begin{eqnarray}\label{inttype3}   
& &\textrm{F.T}\, \left[ 
 \frac{x^2y^2}{(t^2+r^2)^\frac{d+2}{2}} \right],
=\int d^d x(\partial_{p_x} \partial_{p_y})^2 \exp[-i \vec{p}\cdot\vec{r}-i \omega t] \left( \frac{1}{(t^2+r^2)^\frac{d+2}{2}} \right),\\ \nonumber 
&=& \frac{ 2^{ \frac{5 -d}{2} }   \pi^{d-1} }{ \Gamma \left(\frac{d}{2}+1\right)\Gamma \left(\frac{d}{2}-1\right) }
\int dr dt  (\partial_{p_x} \partial_{p_y})^2  ( r^2)^{ \frac{ d-5}{4}} 
\left| \omega \right| ^{\frac{d+1}{2} } K_{\frac{d+1}{2}}\left( {\left| \omega \right|  r} \right)  \, _0\tilde{F}_1\left(;\frac{d-1}{2};-\frac{1}{4} p^2 r^2\right)
  ,\nonumber\\
&=& \frac{\pi ^{d/2} \left( \omega  ^4+ ( p^2 - p_x^2 -p_y^2 ) \left( p^2 - p_x^2 -p_y^2
+2 \omega ^2\right)-p_x^4+6 p_x^2 p_y^2-p_y^4\right)}{\Gamma \left(\frac{d}{2}+1\right)
 \left( \omega^2+p^2 \right)^3}.\nonumber
\end{eqnarray}

 \subsection*{Type 4}
                           In this section we are interested in finding the Fourier transform of the integrals of the type $\partial_\alpha \partial_\beta \frac{1}{(t^2+r^2)^\frac{d-2}{2}}$. 
\begin{eqnarray}\label{inttype4}
\textrm{F.T}\, 
& & \left[ \partial_z\partial_t \frac{1}{(t^2+r^2)^\frac{d-2}{2}} \right]
= \int d^dx \exp[-i \vec{p}\cdot \vec{r}-i \omega t] (-p_z \omega) \frac{1}{(t^2+r^2)^\frac{d-2}{2}},\nonumber\\
&=& (-p_z \omega) \frac{ 2 \pi^{\frac{d-1}{2}} }{\Gamma \left(\frac{d}{2}-1\right)}
\int dr dt r^{d-2}  
 e^{-i t \omega } \left(r^2+t^2\right)^{1-\frac{d}{2}}\, _0\tilde{F}_1\left(;\frac{d-1}{2};-\frac{1}{4} p^2 r^2\right),\nonumber\\
&=& (-p_z \omega)\frac{4 \pi ^{d/2}}{\Gamma \left(\frac{d}{2}-1\right) \left( \omega^2+p^2\right)}.
\end{eqnarray}                                                     

\subsection*{Type 5}
In this section we will evaluate integral of the type $\frac{s_\alpha s_\beta}{s^d}$
 \begin{eqnarray}\label{inttype6}
& & \textrm{F.T}\, \left[ \frac{xt}{(t^2+r^2)^\frac{d}{2}} \right] 
= \int d^d x (i\partial_{p_x})\exp[-i \vec{p}\cdot \vec{r}-i \omega t] 
\left (\frac{t}{(t^2+r^2)^\frac{d}{2}} \right),\nonumber\\
&=&  2 \pi^{\frac{d-1}{2}} 
 \int dr (i\partial_{p_x}) \int dt t e^{-i t \omega }   r^{-2+d} \left(r^2+t^2\right)^{-d/2}\, _0\tilde{F}_1\left(;\frac{d-1}{2};-\frac{1}{4} p^2 r^2\right),\nonumber\\
&=& -\frac{4 \pi ^{d/2} p_x \omega }{\left(p_i^2+p_x^2+p_y^2+p_z^2+\omega^2\right)^2 \Gamma\left[\frac{d}{2}\right]}.
\end{eqnarray}
  \begin{eqnarray}
 \textrm{F.T}\, \left[ \frac{zt}{(t^2+r^2)^\frac{d}{2}} \right] &=&-\frac{4 \pi ^{d/2} p_z \omega }{\left(p_i^2+p_x^2+p_y^2+p_z^2+\omega^2\right)^2 \Gamma\left[\frac{d}{2}\right]}.
\end{eqnarray}
                   
 \subsection*{Type 6}                        
 
\begin{eqnarray}\label{inttype7}  
& &\textrm{F.T}\, \left[ \frac{x^2tz}{(t^2+r^2)^\frac{d+2}{2}}
\right] =\int d^d x(-i\partial_{p_x}^2 \partial_{p_z}) \exp[-i \vec{p}\cdot \vec{r}-i \omega t] 
\left(\frac{t}{(t^2+r^2)^\frac{d+2}{2}}\right),\nonumber\\
&=&2 \pi^{\frac{d-1}{2}} 
\int dr(-i\partial_{p_x}^2 \partial_{p_z}) dt  e^{-i t \omega }  r^{-2+d} t \left(r^2+t^2\right)^{-1-\frac{d}{2}}\, _0\tilde{F}_1\left(;\frac{d-1}{2};-\frac{1}{4} p^2 r^2\right),\nonumber\\
&=& -\frac{2 \pi ^{d/2} p_z \omega  \left(p^2-4 p_x^2 +\omega^2\right)}
{\left(p^2+\omega^2\right)^3 \Gamma\left[1+\frac{d}{2}\right]} .
\end{eqnarray}                                   
\newcommand{\tdelslash}{\tilde{\displaystyle{\not} \partial}}
\newcommand{\Dbslash}{\tilde{\displaystyle{\not} D_b}}
\newcommand{\Dslash}{\tilde{\displaystyle{\not} D}}
\newcommand{\Chi}{\Xi}
\newcommand{\Tr}{\text{tr~}}
\newcommand{\tr}{\text{tr~}}
\newcommand{\rd}{\overrightarrow{\partial}}
\newcommand{\ld}{\overleftarrow{\partial}}
\newcommand{\rD}{\overrightarrow{D}}
\newcommand{\lD}{\overleftarrow{D}}

\section{Evaluating  $\langle T T T \rangle$ in CS vector models}
\label{sec:intro}

$U(N)$ or $O(N)$ Chern-Simons (CS) gauge fields coupled to matter 
in the fundamental representation \cite{Giombi:2011kc,Aharony:2011jz} are a class of 3-dimensional conformal field theories that are exactly solvable in the large $N$ limit. Let us focus our attention on 
theories where the matter is a single fundamental fermion or a single fundamental boson, which are the 
most well-studied.

That these non-supersymmetric field theories are conformal follows from the fact that the 
Chern-Simons level $k$ must be quantized to integer or half-integer values depending on the theory, and therefore 
cannot run. This means that, although the 't Hooft coupling, 
$\lambda=\frac{N}{k}$ is an effectively continuous parameter in the large $N$ limit, it also cannot run. 
CS theory coupled fundamental fermions contains no other adjustable classically relevant or marginal 
couplings that can run (other than a mass  term for the fermion, which can always tuned to zero in 
perturbation theory) and is therefore conformal in perturbation theory. CS theory coupled to fundamental 
bosons has the possibility of a classical $\phi^6$ coupling, however, because this is a triple-trace 
interaction $(\phi^\dagger \phi)^3$ it must also be conformal to all orders in 
$\lambda$ in the large $N$ limit, as one can check in perturbation theory. 

One can also couple CS fields to critical bosons, or ``critical'' 
fermions (i.e., the Gross-Neveu model). 
We then find a bosonization duality where CS gauge theory at small 
$\lambda$ coupled to non-critical fermions, is equivalent to CS theory at large 
$\lambda$ coupled to critical bosons; and vice-versa for critical fermions, described in \cite{Aharony:2012nh}. 

In the large $N$ limit, the planar $\langle T T T \rangle$ correlation function 
to all orders in $\lambda$ was shown to be uniquely determined 
by the slightly-broken higher spin symmetry of the theory in \cite{MZ}. 
In principle, it is also possible to directly calculate the planar correlator to all orders in 
$\lambda$ by summing up the planar Feynman diagrams in lightcone gauge, 
following \cite{Aharony:2012nh, GurAri:2012is}, though to our knowledge that calculation 
has not yet explicitly appeared in the literature.

\subsection{Analysis based on slightly-broken higher spin symmetry}
Let us briefly review 
how the calculation of \cite{MZ} proceeds, for the case of CS theory coupled to 
fundamental fermions.  

We define ``single-trace" operators in the theory to be those
operators obtained by contracting a single fundamental index with an anti-fundamental index. The single-trace primary operators  
consist of a scalar $\tilde{j}_0 \sim \bar{\psi}{\psi}$, with scaling dimension 2, and an 
infinite tower of twist-one operators, that take the schematic form 
$j_s \sim \bar{\psi}\gamma \partial^{s-1} \psi$, one for each spin $s \geq 1$. Following \cite{MZ}, we restrict our attention further to theories containing only even spin currents, 
i.e., CS theory with $O(N)$ gauge group.  (Because of the 
many indices involved it is convenient adopt the convention that all free indices are in a particular null direction
direction, so $j_4 \equiv (j_4)_{----}$, in lightcone coordinates.)

We assume our two-point functions are normalized so 
$\langle j_s j_s \rangle \sim \tilde{N}$, where $\tilde{N}$ is a large parameter proportional to $N$.

In the free theory, all the higher spin currents are conserved. In the interacting theory, 
the divergence of the higher spin currents is restricted by conformal invariance and representation theory. In particular, the divergence of $j_4$, $\partial \cdot j_4 = \partial_\mu (j_4)^\mu{}_{---}$ must take the following form
\begin{equation}
\partial \cdot j_{4} = a_1 \left( (\partial_- \tilde{j}_0) j_2 -\frac{2}{5} \tilde{j}_0 \partial_- j_2 \right). \label{div4}
\end{equation}
where $a_1 \sim \tilde{\lambda}/\tilde{N}$ can be thought of as defining a 
coupling constant.  

Inserting $\partial \cdot j_4$ into a three point function and integrating around each operator gives,
\begin{eqnarray} \label{anom-cons}
& & \int d^3x  \langle  \partial \cdot j_4(x) j_{s_1}(x_1) j_{s_2}(x_2) j_{s_3}(x_3) \rangle  =  \langle [Q_4, j_{s_1}(x_1) j_{s_2}(x_2) j_{s_3}(x_3)] \rangle  \\ & = & \langle [Q_4, j_{s_1}(x_1)] j_{s_2}(x_2) j_{s_3}(x_3) \rangle + \langle  j_{s_1}(x_1) [Q_4,j_{s_2}(x_2)] j_{s_3}(x_3) \rangle + \langle  j_{s_1}(x_1) j_{s_2}(x_2) [Q_4,j_{s_3}(x_3)] \rangle. \nonumber
\end{eqnarray}
where $Q_4 \sim \int d^2x j_4$ is the conserved charge associated with the (almost) conserved current $j_4$. 

The action of the almost conserved charge on $j_s$ is restricted conformal invariance to be of the form,
\begin{equation} \label{Q-action}
[Q_4,j_s] = \sum_{s'=0}^{s+3}c_{s,s'} \partial^{s-s'+3}j_{s'},
\end{equation}
where the coefficients $c_{{s,s'}}$ are a priori unknown constants. Inserting equation \eqref{Q-action} into the 
RHS of equation \eqref{anom-cons} gives a sum of three point functions.

Conformal invariance also restricts the three-point function of the currents in three-dimensions to be of the form:
\begin{equation}
\begin{split}
& \langle j_{s_1}(x_1) j_{s_2}(x_2) j_{s_3}(x_3) \rangle \\ &= \alpha_{s_1,s_2,s_3} \langle j_{s_1}(x_1) j_{s_2}(x_2) j_{s_3}(x_3) \rangle_{\rm free~boson}+\beta_{s_1,s_2,s_3} \langle j_{s_1}(x_1) j_{s_2}(x_2) j_{s_3}(x_3) \rangle_{\rm free~fermion} \\ &\phantom{=} +\gamma_{s_1,s_2,s_3}\langle j_{s_1}(x_1) j_{s_2}(x_2) j_{s_3}(x_3) \rangle_{\rm parity~odd},
\end{split}
\end{equation}
where the $\alpha_{s_1,s_2,s_3}$, $\beta_{s_1,s_2,s_3}$ and $\gamma_{s_1,s_2,s_3}$ are 
unknown coefficients that depend on $\tilde{\lambda}$. Here the subscript ``free fermion'' 
denotes the three point function in the theory of a single real Majorana fermion, the subscript 
``free boson'' denotes the three point function in the theory of a single real boson, 
and ``parity odd'' 
denotes a parity-odd structure that is unique to three dimensions \cite{Giombi:2011rz}, (and may not be exactly conserved). Two-point functions must be of 
the form $\langle j_{s}(x) j_{s}(0) \rangle = \tilde{N}n_{s} \frac{x_-^{2s}}{x^{2s+2}}$, where
$n_s$ are also unknown constants. 

 We thus have several unknown constants: $c_{s,s'}$, $\alpha_{s_1,s_2,s_3}$, 
 $\beta_{s_1,s_2,s_3}$, $\gamma_{s_1,s_2,s_3}$, and $n_{s}$. We can fix some of these 
 unknown constants by choosing a convention for the normalization of the currents. To determine the remaining unknown constants, \cite{MZ} observes that we 
can also write the LHS of the first line of equation \eqref{anom-cons} as follows using equation \eqref{div4}:
\begin{equation}\label{three-2} \begin{split} 
  & \langle \partial \cdot j_4(x) j_{s_1}(x_1) j_{s_2}(x_2) j_{s_3}(x_3) \rangle \\ & = 
a_1 \langle \left( \partial_- \tilde{j}_0 j_2 -\frac{2}{5} \tilde{j}_0 \partial_- j_2 \right)\ j_{s_1}(x_1) j_{s_2}(x_2) j_{s_3}(x_3) \rangle. \end{split}
\end{equation}
Each term on the RHS of this equation factorizes  in the large $\tilde{N}$ limit.  For example, if $s_1=s_2=s_3=2$, this equation becomes:
\begin{equation}\begin{split}
 & \langle \partial \cdot j_4(x) j_{2}(x_1) j_{2}(x_2) j_{2}(x_3) \rangle \\ & = 
a_1  \left( \partial_- \langle \tilde{j}_0(x) j_{2}(x_1) j_{2}(x_2) \rangle \langle j_2(x) j_{2}(x_3) \rangle -\frac{2}{5} \langle \tilde{j}_0(x) j_{2}(x_1) j_{2}(x_2) \rangle  \partial_- \langle j_2(x) j_{2}(x_3) \rangle \right) \\ & + \text{permutations of $x_1$, $x_2$, $x_3$}. \end{split}
\end{equation}
 Inserting equation \eqref{three-2} on the LHS of \eqref{anom-cons}, and using the known expressions for the three-allowed forms for the conformally invariant three-point functions, we can obtain an infinite number of equations (for each choice of spins, we get several equations, roughly one for each choice of points $x_1, x_2, x_3$, since the action of $J_4$ does not commute with conformal transformations) relating the unknown constants listed above. When $s_i \neq 0,2$ the RHS of equation \eqref{three-2} is zero to leading order in $\tilde{N}$, and no integral is required, so solving these equations is relatively straightforward. When one of the spins is zero or $2$, then one must carefully regulate the integral over $x$.
 
 These equations have a two-parameter family of solutions, which are denoted by $\tilde{\lambda}$ and $\tilde{N}$ in \cite{MZ}, and one can determine the coefficients of three point functions, in particular $\alpha_{222}$ and $\beta_{222}$ to all orders in $\tilde{\lambda}$ and leading order in $1/\tilde{N}$. 

\subsection{Summary of results}
Using these results and translating the parameters $\tilde{\lambda}$ and $\tilde{N}$ to $N$ and $k$, with $\lambda=\frac{N}{k}$, the three point function of the stress tensor in $U(N)$ Chern-Simons theory coupled to fundamental Dirac fermions, in terms of $\lambda$ defined using dimensional reduction regularization so that $|\lambda_f|\leq 1$ is:
\begin{equation}
\langle T T T \rangle_\text{int\,  fermion} = n_s(f)  \langle T T T \rangle_\text{free\, boson} + 
n_f(f)  \langle T T T \rangle_\text{free\, fermion} + \gamma(f) \langle T T T \rangle_\text{parity\, odd},
\end{equation}
with 
\begin{eqnarray}
n_s(f)  & = & 2 N \frac{\sin \theta}{\theta} \sin^2 \left( \theta/2 \right),\\
n_f(f) & = & 2 N \frac{\sin \theta}{\theta} \cos^2 \left( \theta/2 \right),\\
\gamma(f) & = & N \frac{\sin^2 \theta}{\theta}.
\end{eqnarray}
where $\theta=\pi \lambda$.

In the $U(N)$ theory coupled to fundamental bosons (non-critical), with 't Hooft coupling $\lambda$ the three point function is:
\begin{equation}
\langle T T T \rangle_\text{int\, boson} = n_s(b) \langle T T T \rangle_\text{free\, boson} + 
n_f(b)  \langle T T T \rangle_\text{free\, fermion} + \gamma(b) \langle T T T \rangle_\text{parity\, odd},
\end{equation}
with 
\begin{eqnarray}
n_s(b) & = & 2 N \frac{\sin \theta}{\theta} \cos^2 \left( \theta/2 \right),\\
n_f(b)  & = & 2 N \frac{\sin \theta}{\theta} \sin^2 \left( \theta/2 \right),\\ 
\gamma(b)  & = & N \frac{\sin^2 \theta}{\theta} .
\end{eqnarray}
where $\theta=\pi \lambda$.

The two point function of the stress tensor in both theories is
\begin{equation}
\langle TT \rangle = 2 N \frac{\sin \theta}{\theta} \langle TT \rangle_\text{free boson}.
\end{equation}
and  $\langle TT \rangle_\text{free boson}=\langle TT \rangle_\text{free fermion}$.

\providecommand{\href}[2]{#2}\begingroup\raggedright\endgroup

\end{document}